# Low-Ripple Modulation Strategy for a Photovoltaic-Based Triple-Port Hydrogen Production System

Shiqi Zhang, Ziang Jiao, Jiaxin Su, Ning Wang, Zheng Li, Xiaoqiang Guo, and Changchun Hua

**Accepted manuscript**

# Low-Ripple Modulation Strategy for a Photovoltaic-Based Triple-Port Hydrogen Production System

Shiqi Zhang, *Graduate Student member, IEEE*, Ziang Jiao, Jiaxin Su, Ning Wang
Zheng Li, and Xiaoqiang Guo, *Senior Member, IEEE,* Changchun Hua, *Fellow, IEEE*

***Abstract*—Among various production methods, hydrogen generation via electrolysis powered by renewable energy plays a key role in achieving large-scale green hydrogen production. The triple active bridge isolated DC-DC conversion system exhibits significant application potential in hydrogen production due to its advantages, such as high energy density, wide step-down ratio, and high reliability. However, the output current ripple at the hydrogen production port critically affects the efficiency of the electrolyzer and the hydrogen production rate. Existing studies have limited optimization effects on current ripple and struggle to achieve dynamic optimization, leading to constrained ripple suppression under dynamic operating conditions. To address this issue, this paper proposes a low-ripple modulation strategy based on coordinated optimization of inner and outer phase-shift angles for multi-port power conversion systems in renewable energy hydrogen production. By establishing an accurate mathematical model, the optimal phase-shift angle combination under minimal current ripple conditions is derived. An improved differential evolution algorithm with adaptive parameter strategy is employed to achieve global optimization under dynamic conditions. Simulation and experimental results demonstrate that the proposed strategy effectively suppresses current ripple, providing an efficient and reliable solution for hydrogen production from fluctuating renewable energy sources.**

***Index Terms*—Hydrogen Production, Low-Ripple Modulation Strategy, Multi-port converter.**

## I. Introduction

Global population growth and technological progress have led to a surge in demand for fossil fuels, but this is unsustainable and causes serious pollution [1]. Renewable energy has become an ideal choice because of its environmental protection, low emissions, and continuous improvement in energy conversion efficiency [2]. Hydrogen energy plays a key role in the transition from fossil fuels to clean energy [3]. It offers advantages such as diverse production methods, high combustion heat value [4], non-polluting combustion products, and energy storage capability [5]. Water electrolysis is the cleanest way to produce hydrogen and works well with renewable energy [6,7].

In a hydrogen production system, the triple-port DC-DC converter stabilizes voltage and current to enhance power quality and ensure efficient operation [8-10]. It performs well under varying load conditions, especially in electrolyzer systems, and supports high voltage conversion, enabling integration with high-voltage DC buses [11,12].

Existing research primarily focuses on optimizing the energy flow, efficiency, and circuit structure of Dual Active Bridge (DAB) converters. A coordinated triple-phase-shift (TPS) control strategy has been developed for an isolated DAB converter, effectively eliminating bidirectional circulating current, achieving full soft-switching control, reducing current stress, and enhancing efficiency [13]. To simplify traditional TPS control with three variables, an automatic TPS DAB converter was developed. It uses smart algorithms to track and predict system states, reducing power loss and improving efficiency [14]. Further, an optimized TPS controlled LCL-DAB converter reduces conduction losses under zero-voltage switching (ZVS) [15].

For hydrogen production, the current ripple is a critical factor. Excessive current ripple can lead to unstable electrolysis reactions, increased energy losses, and reduced hydrogen production [16, 17]. Current ripple has been shown to decrease the hydrogen production efficiency [18], Faraday efficiency [19], and lifetime of electrolyzers [20]. Its impact on hydrogen production efficiency is primarily reflected in changes to the electrolysis voltage and electrolysis current. Therefore, suppressing current ripple is particularly important for water electrolysis-based hydrogen production systems.

For DAB converters, a dual-phase-shift (DPS) control strategy has been proposed for isolated DAB converters to effectively suppress surge current, enhance system efficiency, and reduce output current ripple [21]. Based on this idea, a DPS control method with bidirectional internal phase shift was developed, allowing wider modulation range, lower reactive power, and reduced current ripple compared to traditional methods [22]. Furthermore, an extended phase-shift (EPS)-based EPS-PT switching control technique has been introduced, wherein the switching process is discretized, and the internal and external phase-shift angles are determined according to the input voltage and output current levels [23]. This method offers rapid dynamic response under fluctuating conditions, suppresses voltage ripple, and improves overall power quality.

The isolated triple-port converter is important in renewable hydrogen systems, offering flexible energy control and improved efficiency. However, research on output current ripple optimization for isolated triple-port converters remains limited. Existing studies have proposed an IPS modulation scheme for a triple active bridge (TAB) circuit to reduce ripple by introducing an inner phase-shift angle at the hydrogen production port, achieving optimal ripple suppression [24]. Although this method reduces ripple, its optimization results still require further improvement and are only applicable to a single operating point. In actual renewable energy hydrogen production systems, power generation fluctuations cause the electrolyzer to often operate under dynamic conditions. As a result, the proposed scheme involves extensive computational complexity and fails to achieve continuous current ripple optimization under varying conditions, limiting its applicability in real-world scenarios.

In order to solve the above problems, this paper proposes a low-ripple modulation strategy based on inner and outer phase-shift angle optimization (IEPSM). An IEPSM strategy is proposed with analytical modeling to optimize the output current ripple at the electrolyzer port. An improved adaptive DE algorithm is employed to optimize the phase-shift angles, achieving faster convergence and higher accuracy. The proposed strategy adapts to continuous current ripple regulation under different operating conditions and has been validated through simulations and experiments with a connected electrolyzer load model, demonstrating its effectiveness in practical hydrogen production applications.

The rest of this article is structured as follows. Section II introduces the control principle of the TAB circuit and derives the output circuit expression of the hydrogen production port. Section III proposes an optimization algorithm for ripple suppression, which can achieve continuous ripple regulation under dynamic parameter changes. Section IV verifies the feasibility of the proposed method through simulation. Section V verifies the feasibility of the proposed method through experiments. Finally, Section VI summarizes the whole article.

TABLE I
COMPARISON OF EXISTING RESEARCH

| Ref. | Converter Type | Control Strategy | Main Contribution | Electrolyzer Load Considered |
|---|---|---|---|---|
| [13] | DAB | Triple Phase Shift (TPS) control strategy | Effectively eliminates bidirectional circulating current, achieves full soft-switching, reduces current stress, and improves converter efficiency | No |
| [14] | DAB | Automated TPS DAB converter using intelligent algorithms to track and predict system states | Reduces power losses and enhances converter efficiency | No |
| [15] | DAB | Optimized TPS-controlled LCL-DAB converter based on Fundamental Harmonic Analysis | Reduces conduction loss and ensures Zero Voltage Switching | No |
| [21] | DAB | Dual Phase Shift control strategy | Suppresses surge current, enhances system efficiency, and reduces output current ripple | No |
| [22] | DAB | DPS method with bidirectional internal phase shift | Offers wider modulation range, lower reactive power, and reduced current ripple compared to conventional methods | No |
| [23] | DAB | Extended Phase Shift (EPS)-based EPS-PT switching control technique | Discretizes switching process and sets internal and external phase shift angles based on input voltage and output current levels | No |
| [24] | TAB | Hybrid control combining PWM and phase-shift in TAB circuit | Suppresses current ripple by introducing internal phase shift at the hydrogen production port, achieving partial ripple optimization | No |
| The paper | TAB | IEPSM-based dynamic ripple optimization for electrolyzer load | Achieves lower ripple at the hydrogen production port; dynamic ripple optimization validated with electrolyzer load model | Yes |

## II. OPTIMIZED CONTROL STRATEGIES FOR THE SYSTEMS

### *A. Photovoltaic hydrogen production system composition*

According to the hydrogen production system design requirements, as shown in Fig. 1, Port 1 of the triple-port hydrogen production converter connects to the photovoltaic power generation unit, Port 2 connects to the battery, and Port 3 connects to the electrolyzer unit. The system consists of a photovoltaic module, DC bus, multi-port DC/DC converter, battery module, and electrolyzer module. The photovoltaic module has a maximum power output of 20 kW. The electrolyzer module adopts PEM technology and operates at a rated power of 11.6 kW. In the triple-port hydrogen production converter, the turns ratio of the three-winding transformer is $N_1$: $N_2$: $N_3$, and each full-bridge unit is composed of switching devices $S_{x1}$~$S_{x4}$(x=a, b, c).

The hydrogen production technology adopted in this paper is Proton Exchange Membrane (PEM) electrolysis. This type of electrolyzer offers advantages such as fast dynamic response and strong adaptability to the fluctuations of renewable energy sources, making it more suitable for water electrolysis driven by variable renewable energies like wind and solar [25]. The dynamic model of the PEM electrolyzer constructed in this study is based on its electrochemical characteristics. Since the cathodic reaction proceeds much faster than the anodic reaction during actual operation, the model primarily focuses on the cathodic process [26].

Fig. 2 shows the equivalent circuit model of the PEM electrolyzer, which consists of the following components: 1) Two resistors ($R_a$, $R_c$)–capacitor ($C_a$, $C_c$) branches used to model the dynamic characteristics, Gibbs energy, and losses of the anode and cathode. 2) A resistor ($R_{mem}$) is used to represent the ohmic overvoltage and proton membrane losses. 3) A DC voltage source ($E_{rev}$) is used to simulate the reversible voltage. The model represents a minimum unit. The total power of the electrolyzer can be adjusted by varying the number of these minimum units.

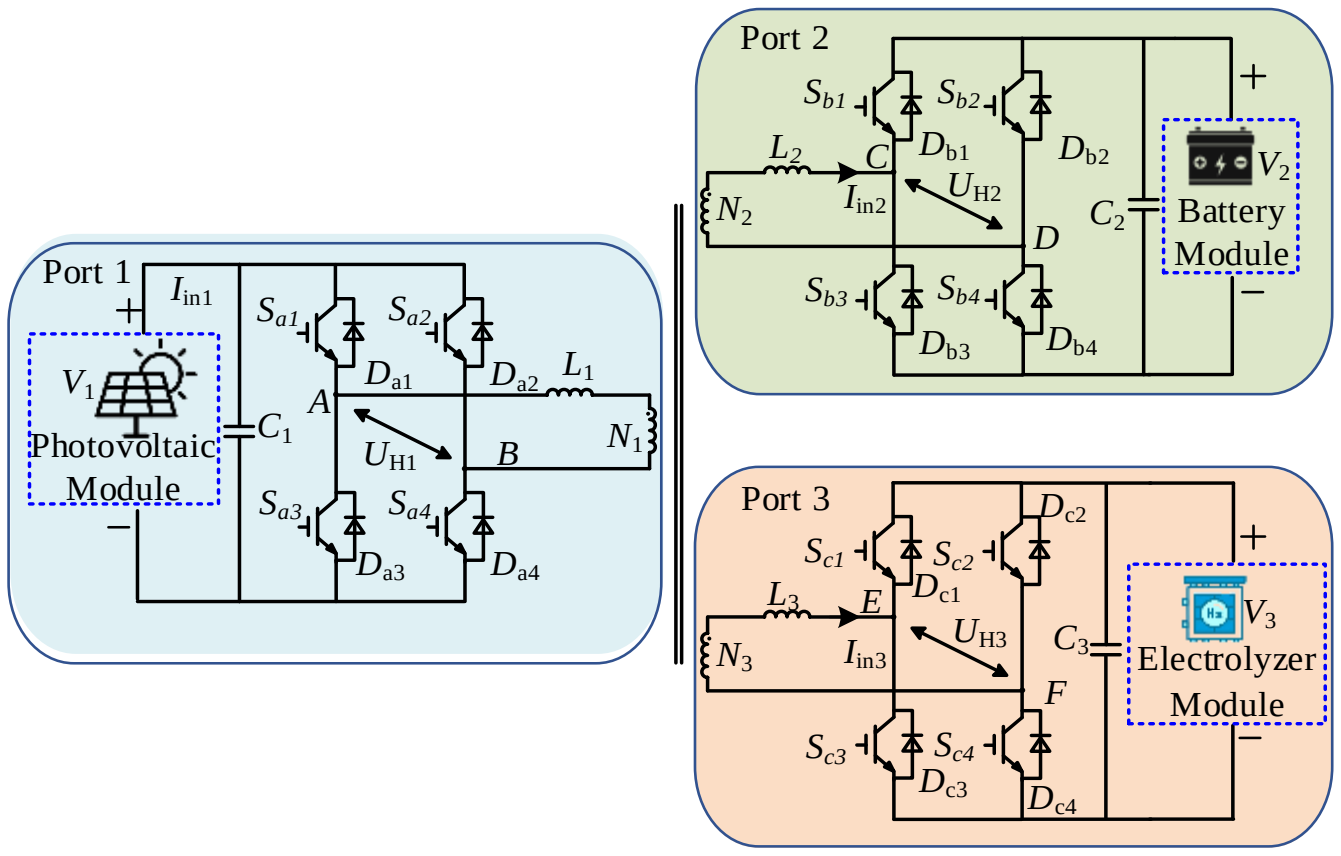

Fig. 1. Triple-Port Hydrogen Production Conversion System Topology.

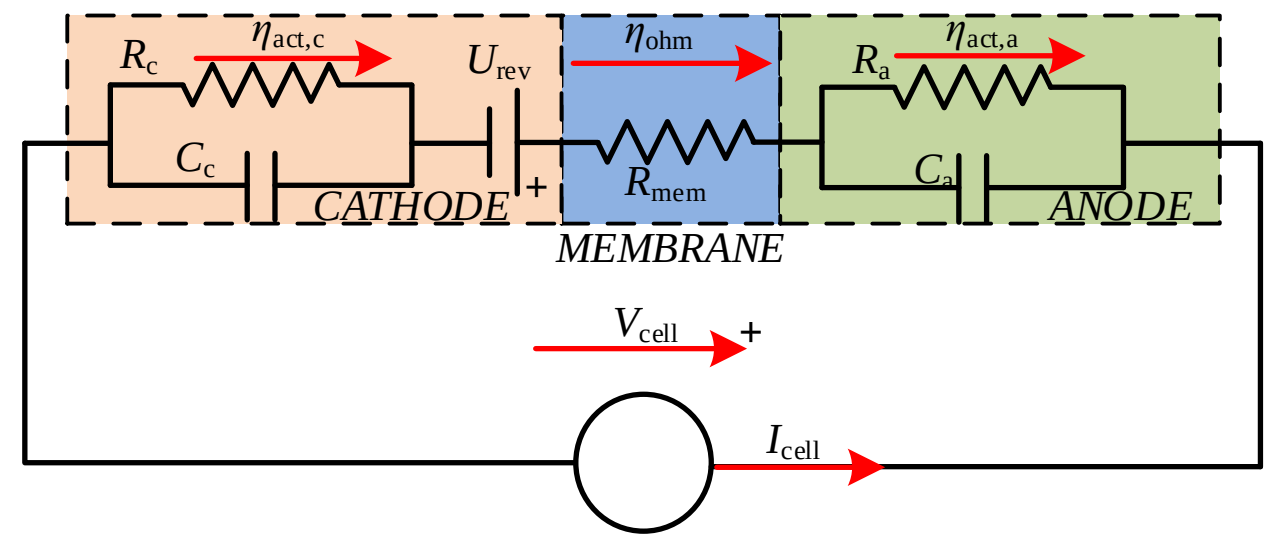

Fig. 2. Equivalent Circuit Model of the PEM Electrolyzer.

## B. Equivalent circuit of photovoltaic hydrogen production triple-port power conversion system

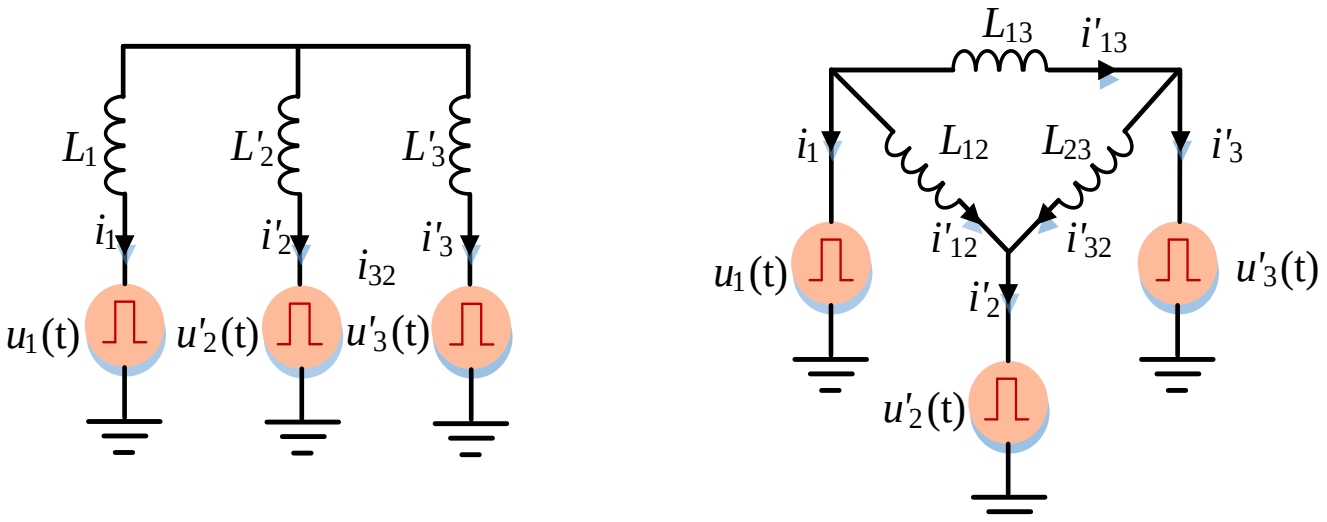

Fig. 3. TAB equivalent circuit diagram.

For easier analysis, the voltages and currents of ports 2 and 3 are referred to as port 1. The TAB circuit is simplified to a Y-type circuit as shown in Fig. 3(a). Here, $u_2$ and $u_3$ are the original port voltages, while $u_2'$ and $u_3'$ are the converted voltages. Assume transformer ratios are $N_1$: $N_2$: $N_3$= $n_1$:$n_2$:$n_3$. $L_2'$ and $L_3'$ are the equivalent leakage inductances after conversion. Therefore, the calculated voltage inductance can be expressed as $u_2'= u_2* n_1/ n_2$, $u_3'= u_3* n_1/ n_3$, $L_2'= L_2* n_1/ n_2$, $L_3'= L_3* n_1/ n_3$.

To simplify the description and calculation process, the Y-circuit is converted to an equivalent delta circuit, as shown in Fig. 2(b) [27]. Then calculate the equivalent inductances $L_{12}$, $L_{13}$, and $L_{23}$ using the given formulas.

$$\begin{cases} L_{12} = L_1 + L_2' + \dfrac{L_1 L_2'}{L_3'} \\ L_{23} = L_2' + L_3' + \dfrac{L_2' L_3'}{L_1} \\ L_{13} = L_1 + L_3' + \dfrac{L_1 L_3'}{L_2'} \end{cases} \tag{1}$$

## C. Analysis of IEPSM Phase Shift Strategy

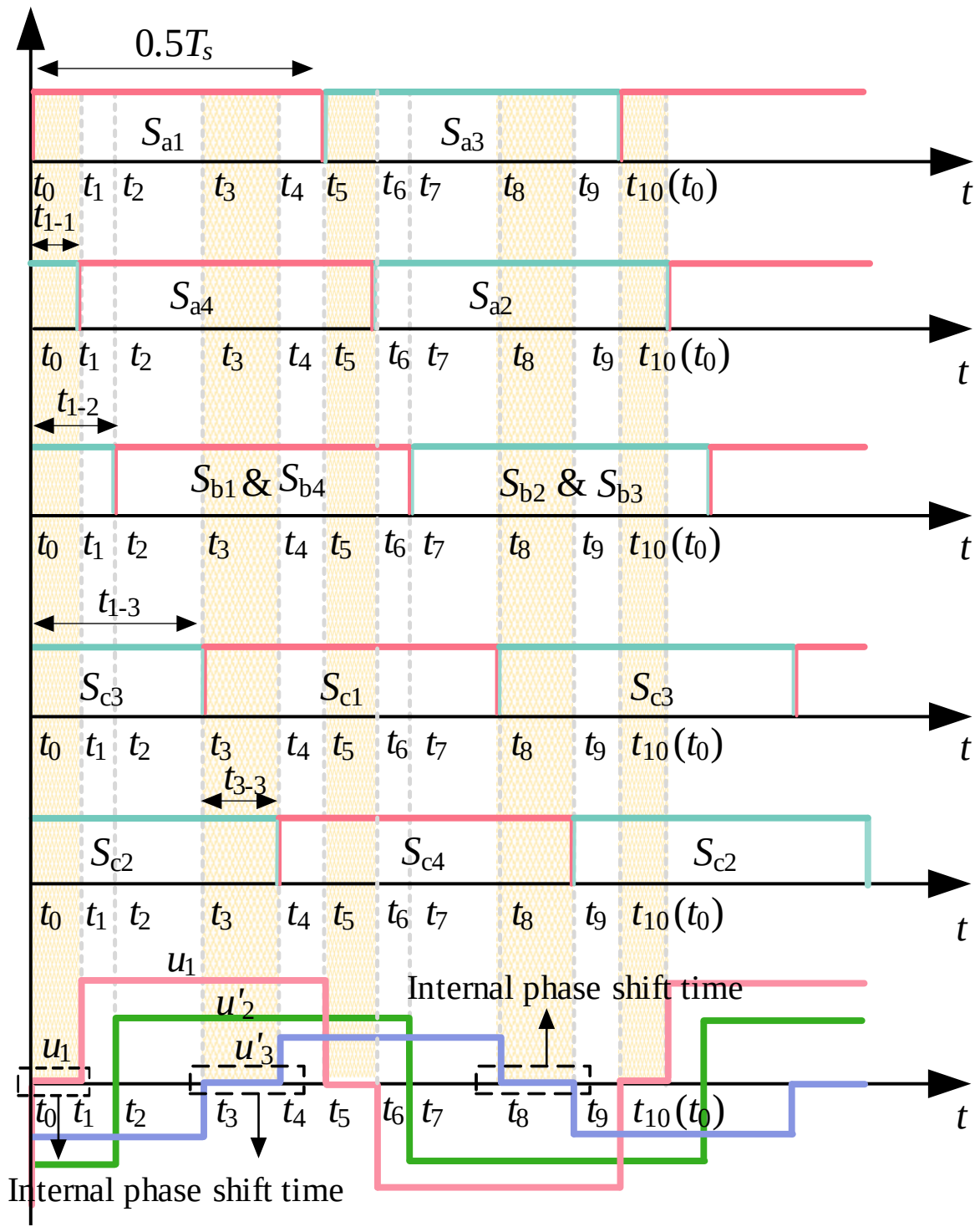

Fig. 4. Low ripple modulation based on IEPSM.

Electrolyzers need a low DC voltage for hydrogen production, usually provided by a DC/DC step-down converter [28]. Current ripple can reduce hydrogen production efficiency [29]. Internal phase-shift modulation offers more control flexibility by adjusting phase shifts, helping to reduce ripple and improve efficiency and reliability.

To reduce load port current ripple, this paper proposes an IEPSM-based low-ripple modulation strategy. By applying internal phase shifts to ports 1 and 3, a three-level square wave voltage is generated, improving output current stability. Proper combinations of internal and external phase shifts help minimize ripple. Fig. 4 shows the port driving signals and load voltage under this strategy.

Here, $t_{1\text{-}1}$ and $t_{3\text{-}3}$ are the phase shift time differences between diagonal switches at the load port. As shown in Fig. 4, after introducing the internal phase shift angle at ports 1 and 3, the voltage waveforms at these ports transition from the rectangular square wave voltage under SPS (Single Phase Shift) control to a three-level AC voltage. The energy flow modes can be divided into two categories: single-input dual-output and dual-input single-output. The working principles of TAB converters under different power flows are very similar. This article only discusses the dual-input single-output mode. With port 1 as the main input, Eq. (2) represents the power relationship among the three ports:

$$\begin{cases} P_1 = -(P_{1\text{-}2}+P_{1\text{-}3}) \\ P_2 = P_{1\text{-}2} - P_{2\text{-}3} \\ P_3 = P_{1\text{-}3} + P_{2\text{-}3} \end{cases} \tag{2}$$

From Eq. (2), it can be inferred that to solve for the power $P_3$ received by the load port, it is necessary to determine $P_{1\text{-}3}$ and $P_{2\text{-}3}$. Taking the solution of $P_{1\text{-}3}$ as an example, the corresponding voltage waveforms are illustrated in Fig. 5:

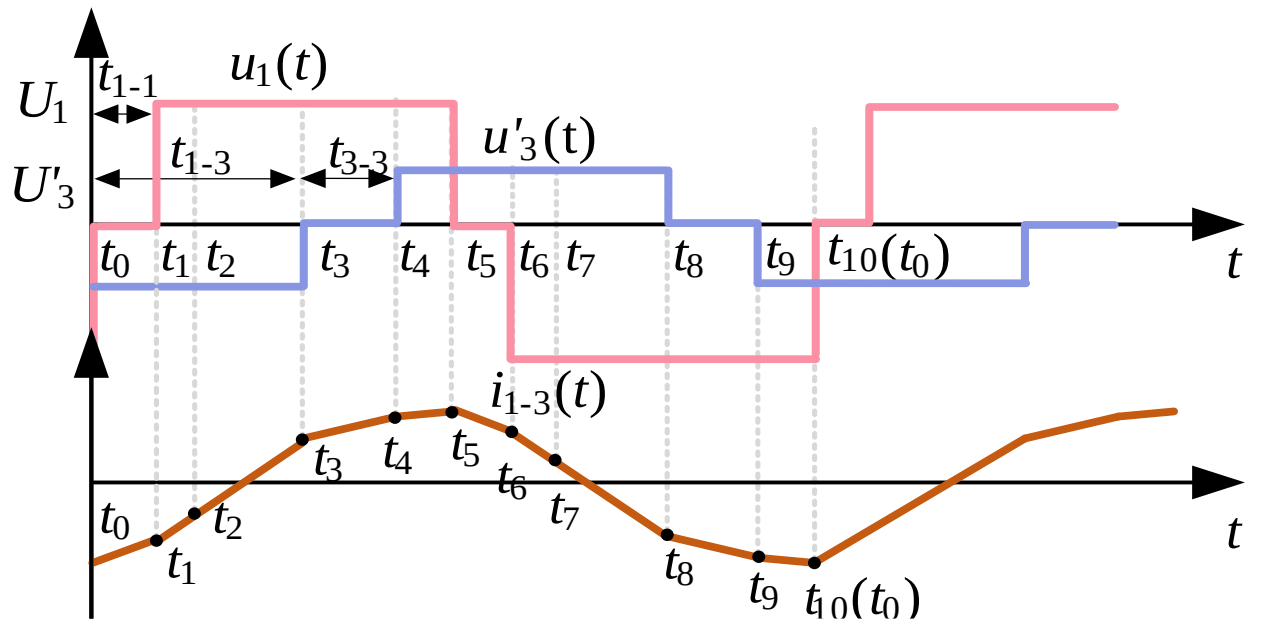

Fig. 5. 1-port and 3-port voltage waveforms.

Based on the equivalent circuit in Fig. 3(b), the relationship between ports 1 and 3 is given by Eq. (3):

$$L_{13}\frac{di_{L13}}{dt} = u_1 - u'_3 \tag{3}$$

By integrating the current between ports 1 and 3 over a full cycle, Eq. (4) gives the average current expression for both ports:

$$i_{L13}(t) = L_{13}\frac{di_{L13}}{dt} = \frac{1}{L_{13}}\int (u_1 - u'_3)dt \tag{4}$$

Under steady-state conditions, the output current and port voltages vary periodically. Within the half-cycle from 0 to T/2, the maximum and minimum values of the waveform are equal in magnitude but opposite in sign, as described by Eq. (5).

$$f(t) = -f(t+T/2) \tag{5}$$

Over the full period from 0 to T, the signal values repeat every T interval, which is expressed accordingly in the following equation.

$$f(t) = f(t+T) \tag{6}$$

Without considering system losses, that is, assuming the input power equals the output power, $P_{1-3}$ is given by Eq. (7):

$$P_{1-3} = \frac{1}{0.5T}\int_0^{T/2} u_1(t) i_{L13}(t)dt \tag{7}$$

$$P_{1-3} = -\frac{u_1 u'_3}{L_{13}T}(-Tt_{1-3} + 2t_{1-3}^2 + 0.5Tt_{1-1} - 2t_{1-1}t_{1-3} \\ + t_{1-1}^2 - 0.5Tt_{3-3} + 2t_{1-3}t_{3-3} - t_{1-1}t_{3-3} + t_{3-3}^2) \tag{8}$$

Similarly, the power expression is transferred from port 2 to port 3, and the output current of the load port can be obtained.

$$P_{2\text{-}3} = \frac{U'_2 U'_3}{L_{23}T}(Tt_{2\text{-}3} - 2t_{2\text{-}3}^2 + 0.5Tt_{3\text{-}3} - 2t_{2\text{-}3}t_{3\text{-}3} - t_{3\text{-}3}^2) \tag{9}$$

According to the power flow, the output current of port 3 can be solved using Eq. (10).

$$I_{out} = \frac{P_{1\text{-}3}+P_{2\text{-}3}}{U'_3} = \frac{u'_2(Tt_{2\text{-}3} - 2t_{2\text{-}3}^2 + 0.5Tt_{3\text{-}3} - 2t_{2\text{-}3}t_{3\text{-}3} - t_{3\text{-}3}^2)}{L_{23}T} + \\ \frac{u_1(Tt_{1-3} + t_{1-1}t_{3-3} - (t_{1-1} - t_{1-3})^2 - (t_{1-3} + t_{3-3})^2 - 0.5T(t_{1-1} - t_{3-3}))}{L_{13}T} \tag{10}$$

The load's DC output is rectified from the hydrogen port, so the transformer's current waveform directly affects output quality. Therefore, suppressing the peak current at the transformer's hydrogen production port can effectively reduce the output current ripple. Under the IEPSM modulation strategy, the output voltage at the load port is a three-level high-frequency AC voltage, and the specific equivalent circuit is shown in Fig. 3(b). Port current $i'_3$ is influenced by $u_1(t)$, $u'_2(t)$ and $u'_3(t)$, and can be expressed as Eq. (11):

$$i'_3 = i'_{1\text{-}3} - i'_{3\text{-}2} = \frac{u_1(t) - u'_3(t)}{L_{13}} - \frac{u'_3(t) - u'_2(t)}{L_{23}} \tag{11}$$

TABLE II
CURRENT WAVEFORM AT THE LOAD PORT WITHIN HALF A CYCLE

| Current waveform | Current expression | Time period |
|---|---|---|
| | $i'_3(t_1) = (\frac{u'_3}{L_{13}} - \frac{-u'_3 + u'_2}{L_{23}})(t_1 - t_0) + i_0$ | $t_0$~$t_1$ |
| | $i'_3(t_2) = (\frac{u_1 + u'_3}{L_{13}} - \frac{-u'_3 + u'_2}{L_{23}})(t_2 - t_1) + i_1$ | $t_1$~$t_2$ |
| | $i'_3(t_3) = (\frac{u_1 + u'_3}{L_{13}} - \frac{-u'_3 - u'_2}{L_{23}})(t_3 - t_2) + i_2$ | $t_2$~$t_3$ |
| | $i'_3(t_4) = (\frac{u_1}{L_{13}} + \frac{u'_2}{L_{23}})(t_4 - t_3) + i_3$ | $t_3$~$t_4$ |
| | $i'_3(t_5) = (\frac{u_1 - u'_3}{L_{13}} - \frac{u'_3 - u'_2}{L_{23}})(t_5 - t_4) + i_4$ | $t_4$~$t_5$ |

Table II shows the load port current over one cycle. The maximum current occurs at time points $t_0$ and $t_5$. To find its expression, an iterative method is used to calculate the current at $t_5$. Current segments from $t_0$ to $t_5$ follow the relationships in Table II and satisfy Eq. (12).

$$i'_3(t_0) = -i'_3(t_5) \tag{12}$$

Each time period and the relationship between phase shift times satisfy Eq. (13).

Using Eq. (12), Eq. (13) along with Table II, the load port current waveform over one cycle is obtained. The peak current occurs at $t_0$ and $t_5$, when the absolute value of the load current is the largest. This paper uses an iterative method to derive the formula for peak current. The current relationships for each time segment $t_0$ to $t_5$ of $i'_3$ are shown in Table II:

$$\begin{cases} t_1 - t_0 = t_{1\text{-}1} \\ t_2 - t_1 = t_{1\text{-}2} - t_{1\text{-}1} \\ t_3 - t_2 = t_{1\text{-}3} - t_{1\text{-}2} = t_{2-3} \\ t_4 - t_3 = t_{3\text{-}3} \\ t_5 - t_4 = 0.5T - t_{1\text{-}3} - t_{3\text{-}3} \end{cases} \tag{13}$$

$$i_3'(t_0) = -\frac{0.25Tu_1 - 0.5t_{1-1}u_1 + t_{1-3}u_3' + 0.5t_{3-3}u_3' - 0.25Tu_3'}{L_{13}} - \frac{0.25Tu_2' - t_{1-2}u_2' + t_{1-3}u_3' + 0.5t_{3-3}u_3' - 0.25Tu_3'}{L_{23}} \tag{14}$$

Then, by substituting it into the expression for $i'_3(t_5)$, the peak current expression for the load port $i'_3$(peak) can be obtained.

$$i'_{3(peak)} = \frac{0.25Tu_1 - 0.5t_{1-1}u_1 + t_{1-3}u_3' + 0.5t_{3-3}u_3' - 0.25Tu_3'}{L_{13}} + \frac{0.25Tu_2' - t_{1-2}u_2' + t_{1-3}u_3' + 0.5t_{3-3}u_3' - 0.25Tu_3'}{L_{23}} \tag{15}$$

The phase shift time $t_{1\text{-}2}$ of ports 1 and 2 is fixed to 12μs, and the influence of the phase shift time $t_{1\text{-}1}$, $t_{1\text{-}3}$, and $t_{3\text{-}3}$ on $i'_{3(\text{peak})}$ is analyzed.

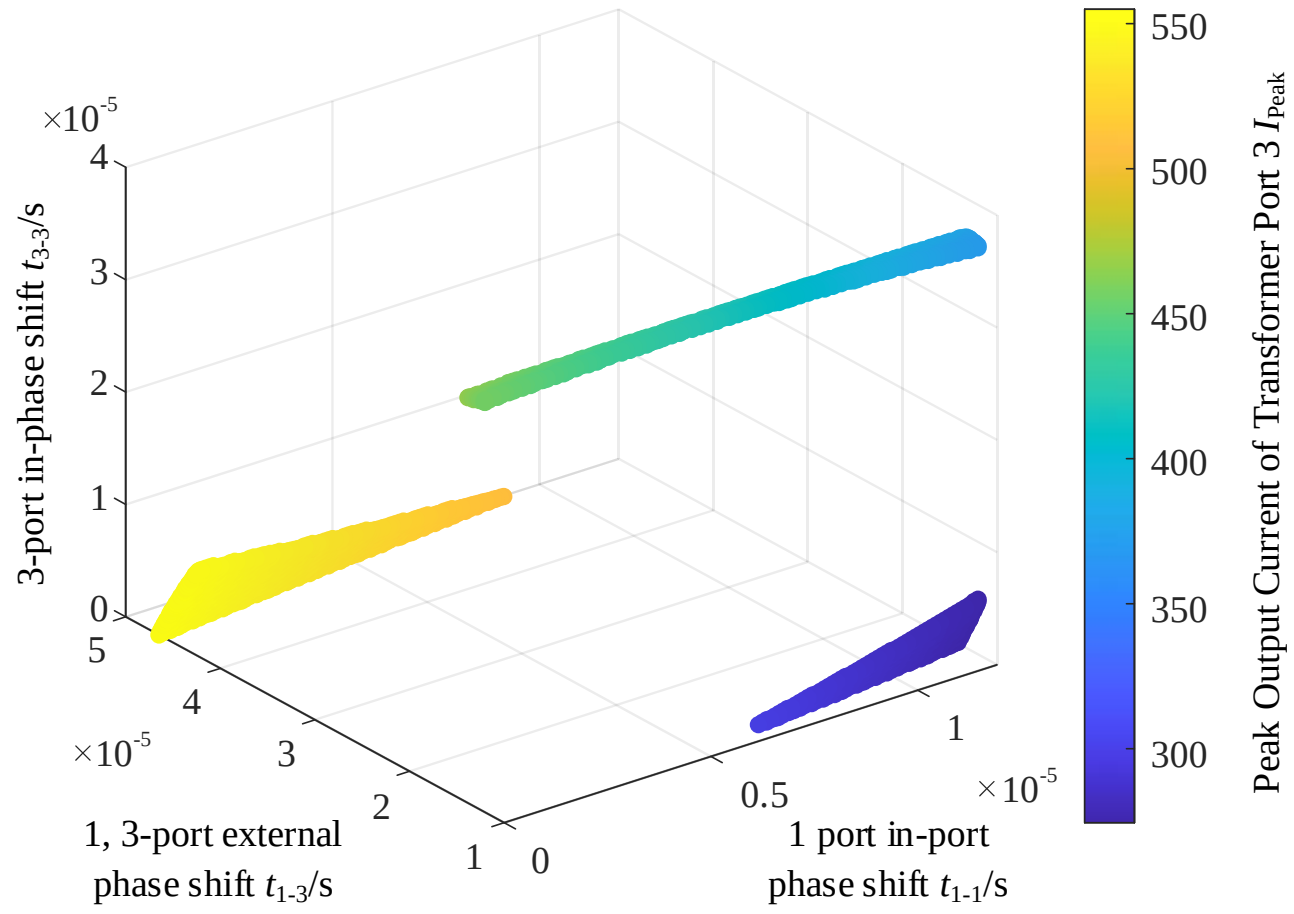

Fig. 6. Internal and external phase shift angles for 100 A output current.

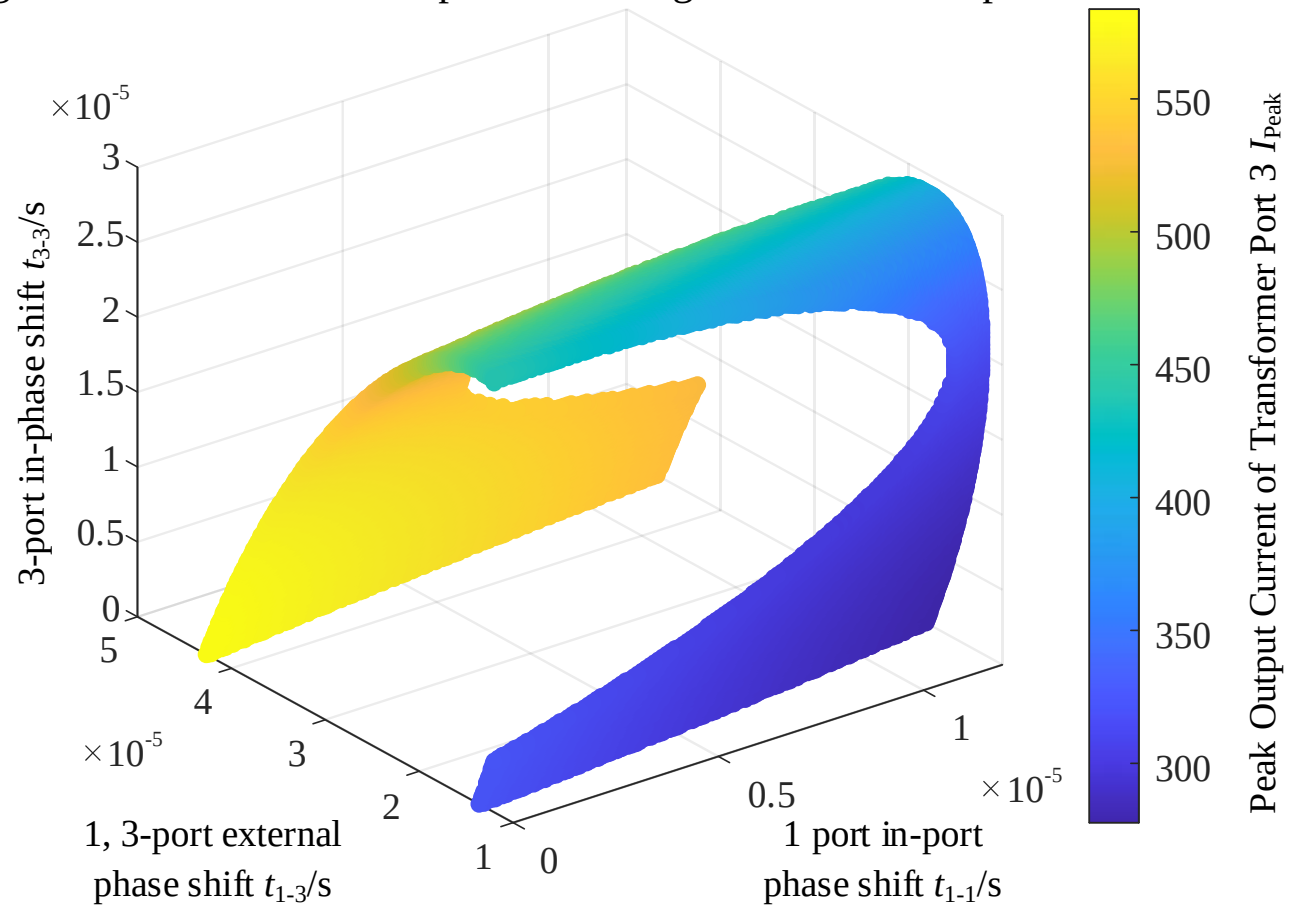

Fig. 7. Internal and external phase shift angles for 150 A output current.

Figs. 6, 7, and 8 show how the peak current at port 3 changes at 100A, 150A, and 200A output settings. The diagram illustrates the impact of different phase-shift angle combinations on $i'_3$(peak), highlighting the strong correlation between modulation parameters and ripple behavior. It provides intuitive insight into how coordinated phase adjustment affects current ripple. For each scenario, the three phase shift variables $t_{1\text{-}1}$, $t_{1\text{-}3}$, and $t_{3\text{-}3}$ are evenly sampled within the range [0, T/2], with $500^3$ points in total across the three dimensions. The corresponding peak current is calculated for each parameter combination. The X-axis represents the internal phase shift time $t_{1\text{-}1}$ at port 1, the Y-axis represents the external phase shift time $t_{1\text{-}3}$ between port 1 and port 3, and the Z-axis represents the internal phase shift time $t_{3\text{-}3}$ at port 3. The color scale reflects the magnitude of the peak current, with hues closer to blue denoting lower peak current values, and those nearer to yellow indicating higher peak current levels.

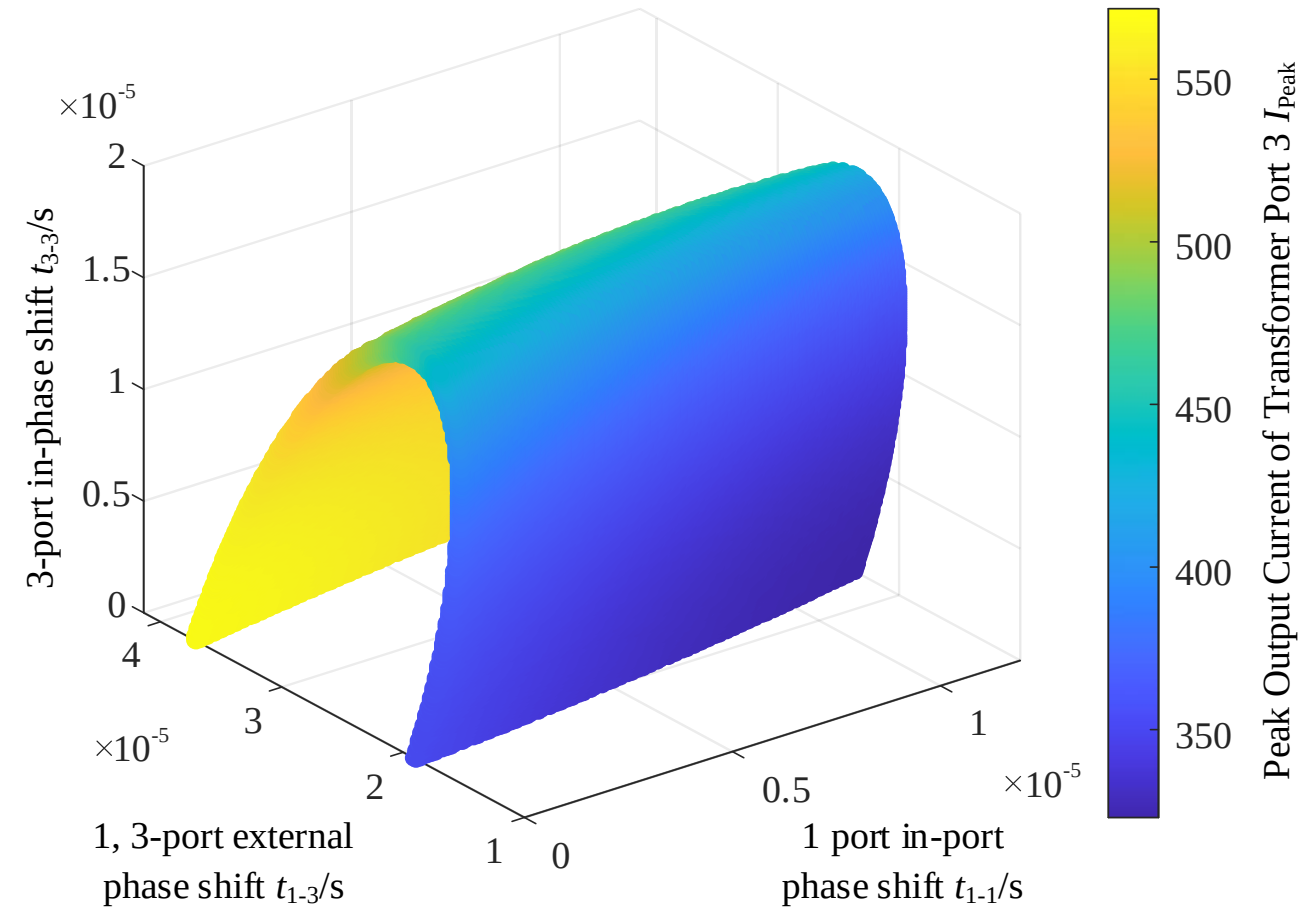

Fig. 8. Internal and external phase shift angles for 200 A output current.

## III. Adaptive Differential Evolution for Optimal Phase-Shift Angle Combination

To minimize the output current ripple of port 3, an improved DE algorithm with adaptive parameters is proposed to optimize the phase shift angles and minimize $I_{\text{peak}}$, as shown in Fig. 9.

Differential Evolution (DE) is a population-based stochastic search method [30,31]. Similar to all other evolutionary algorithms (EA), DE utilizes mutation, crossover, and selection operators in each generation to achieve global optimization, making it one of the most efficient EA currently in use [32].

Since the initial population selection is random and global, the DE algorithm reduces reliance on initial values. However, when the objective function has multiple extreme values, it is easy to fall into a local optimum, and the solution fluctuates widely, which is a limitation of the DE algorithm. To address the DE algorithm has large solution fluctuations and is easy to fall into local optimality in TAB phase shift angle optimization, this paper proposes a fusion strategy that combines the elite retention mechanism with the strengthening constraint mechanism to suppress the fluctuation range of the solution, and introduces an adaptive parameter strategy to enhance the algorithm's ability to jump out of the local optimality.

To minimize the output current ripple at port 3 in the TAB circuit, this paper employs a differential evolution algorithm for optimization. The objective function is to minimize the peak value of the output port current, as expressed in Eq. (16). The constraints include power balance and constant output current, as represented in Eq. (17).

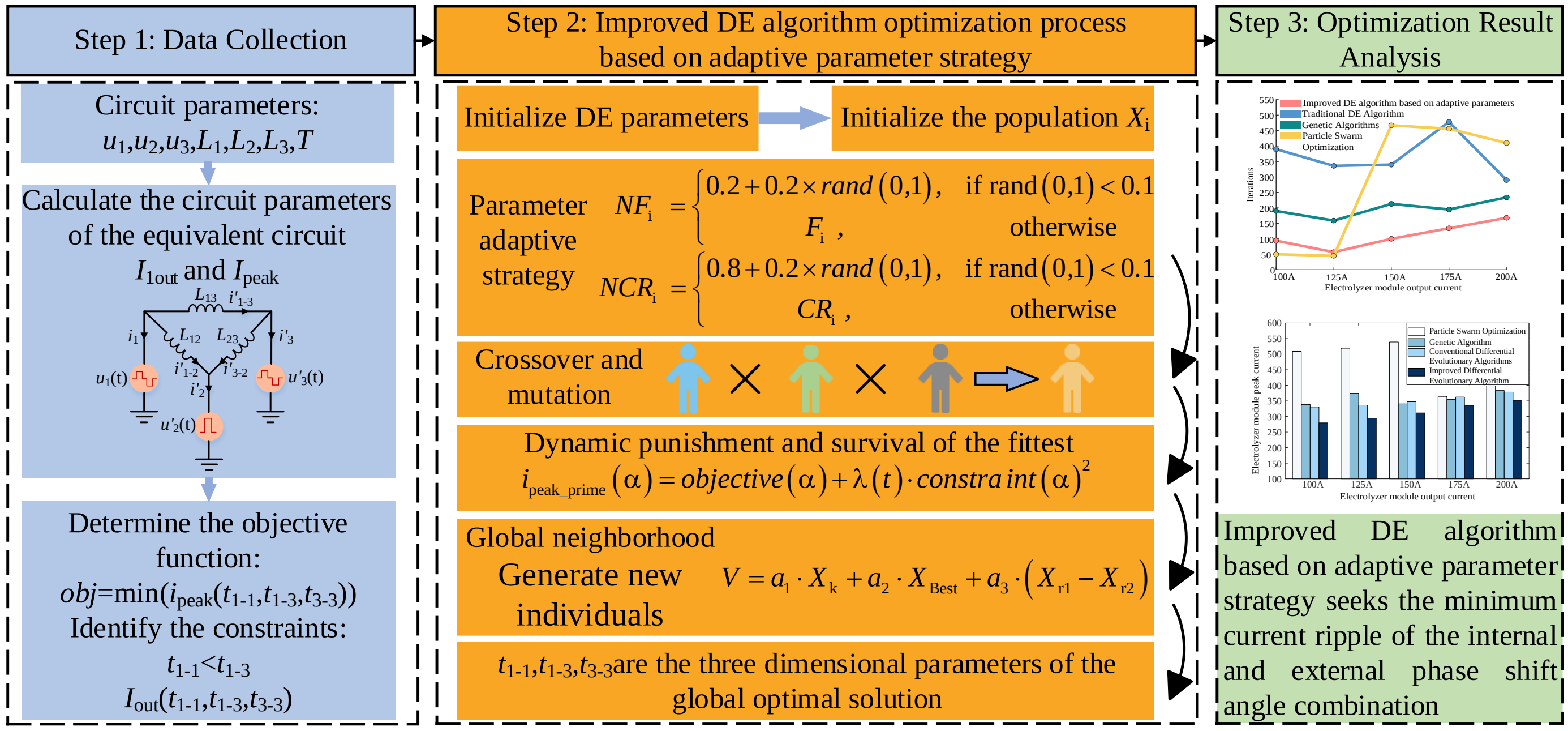

Fig. 9. Improved differential evolution algorithm with adaptive parameters for ripple optimization.

$$objective = \min\ i_{peak}\left(t_{1-1}, t_{1-3}, t_{3-3}\right) \tag{16}$$

$$\begin{cases} P_{in} - P_{out} - P_{loss} = 0 \\ I_{out} = \text{constant value} \end{cases} \tag{17}$$

Here, $I_{peak}$ denotes the peak transformer current at the hydrogen production port.

*A. Improved differential evolution algorithm*

Elite retention prevents the loss of high-quality solutions during the iteration process by retaining the historical optimal solution. However, excessive reliance on elites may lead to a decline in population diversity. This paper combines the elite retention strategy with a global neighborhood search strategy to avoid the algorithm falling into a specific local optimum while ensuring that the optimization direction does not deviate from the global optimum.

To address the TAB power balance constraint $P_{in}$-$P_{out}$-$P_{loss}$=0, traditional static penalty function methods require manual adjustment of weight coefficients, which can easily lead to over-penalization or under-penalization. This paper adopts a dynamic penalty function, dynamically integrating the degree of constraint violation into the objective function:

$$i_{peak_prime}\left(\alpha\right) = objective\left(\alpha\right) + \lambda\left(t\right)\cdot constraint\left(\alpha\right)^2 \tag{18}$$

Among them, $\lambda(t)$ increases as the iteration count $t$ grows.

$$\lambda\left(t\right) = \lambda_0 \cdot e^{\frac{t}{T_{max}}} \tag{19}$$

The initial penalty coefficient $\lambda_0$ is set to $10^3$ to allow flexible exploration in early iterations, while the final value $\lambda(T_{max}) = 10^6$ enforces strict constraint satisfaction at convergence. These values are chosen based on empirical tuning and reflect a gradual shift from exploration to exploitation [33]. The dynamic adjustment mechanism allows for some constraint violations in the early stages to explore the global space, while enforcing strict constraint satisfaction in the later stages.

The power balance constraint of the TAB exhibits strong nonlinear characteristics. The dynamic penalty function prevents the algorithm from falling into infeasible regions due to strict constraints in the early stages, while ensuring the engineering feasibility of the final solution.

*B. Parameter Adaptation Strategy*

The performance of the DE algorithm is significantly influenced by the scaling factor $F$ and the crossover probability $CR$ [34]. The scaling factor Fi and crossover probability $CR$i are key parameters in the adaptive DE algorithm. Fi controls the mutation amplitude and thus influences the exploration range, while $CR$i determines the recombination rate and affects population diversity. To avoid premature convergence, both parameters are probabilistically regenerated in each generation. Only successful updates are retained, allowing effective parameter values to be adaptively preserved throughout the optimization process [35]. To ensure that the algorithm exhibits strong adaptability and convergence when dealing with complex objective functions featuring multiple extrema, this paper adopts a global neighborhood search-based adaptive differential evolution algorithm. The algorithm adjusts parameters automatically and uses global search to improve individual quality toward the optimum. The adaptive parameter mechanism dynamically balances exploration and exploitation, avoiding the limitations imposed by global parameters on population diversity and reducing solution fluctuations.

In the global neighborhood search-based adaptive differential evolution algorithm, each individual $i$ in every generation has independent $F_i$ and $CR_i$. For individual $i$ in the $n_{th}$ generation population, the scaling factor is $F_i$, and the crossover probability is $CR_i$. $NF_i$ and $NCR_i$ are the scaling factor and crossover probability of the trial individual $U_i$. For the $n_{th}$ generation, $NF_i$ and $NCR_i$ are modified as follows:

$$NF_i = \begin{cases} 0.2 + 0.2 \times rand\left(0,1\right), & \text{if rand}\left(0,1\right) < 0.1 \\ F_i\ , & \text{otherwise} \end{cases} \tag{20}$$

$$NCR_{\mathrm{i}} = \begin{cases} 0.8 + 0.2 \times rand(0,1), & \text{if } \mathrm{rand}(0,1) < 0.1 \\ CR_{\mathrm{i}}, & \text{otherwise} \end{cases} \quad (21)$$

Eq. (20)regenerates $F_{\mathrm{i}}$ with a 10% probability, ensuring that $F_{\mathrm{i}}$ is distributed within 0.2, 0.4, thereby enhancing local exploitation capabilities. Eq. (21) regenerates $CR_{\mathrm{i}}$ with a 10% probability, ensuring that $CR_{\mathrm{i}}$ is distributed within 0.8, 1.0, thereby improving global exploration efficiency. The modified $NF_{\mathrm{i}}$ and $NCR_{\mathrm{i}}$ are used to generate the trial individual for the ith individual in the $t_{\mathrm{th}}$ generation population. The updated values $F_{\mathrm{inext}}$ and $CR_{\mathrm{inext}}$ for the $i_{\mathrm{th}}$ individual in the $(n+1)_{\mathrm{th}}$ generation population are modified as follows:

$$F_{\mathrm{inext}} = \begin{cases} NF_{\mathrm{i}}, & \text{if } fit(U_{\mathrm{i}}) \text{ superior to } fit(X_{\mathrm{i}}) \\ F_{\mathrm{i}}, & \text{otherwise} \end{cases} \quad (22)$$

$$CR_{\mathrm{inext}} = \begin{cases} NCR_{\mathrm{i}}, & \text{if } fit(U_{\mathrm{i}}) \text{ superior to } fit(X_{\mathrm{i}}) \\ CR_{\mathrm{i}}, & \text{otherwise} \end{cases} \quad (23)$$

The *fit*(X) denotes the fitness evaluation of individual X. It is calculated as the sum of the objective function and the dynamic penalty term associated with constraint violation. This value is used to assess the quality of an individual solution in the population. The Differential Evolution algorithm compares the fitness of the trial and current individuals. If the trial individual's fitness is higher, it replaces the current individual, improving population fitness and advancing the optimization. The values of *F*i and *CR*i are updated only when the generated trial solution is superior to the original solution, thereby retaining successful parameter combinations.

To address the issue of slow convergence in DE, this paper adopts a global neighborhood search strategy to accelerate convergence. This strategy generates new solutions by hybridizing the current optimal solution with random individuals, which are then compared with the original individuals to select the best for survival.

Let $X_{\mathrm{k}}$ be an individual randomly selected from the $n_{\mathrm{th}}$ generation population, $X_{\mathrm{Best}}$ represent the most excellent individual in the current population, and $X_{\mathrm{r1}}$ and $X_{\mathrm{r2}}$ denote two individuals randomly chosen from the current population that are different from $X_{\mathrm{k}}$. By applying the global neighborhood search strategy, a new individual *V* is obtained as follows:

$$V = a_1 \cdot X_{\mathrm{k}} + a_2 \cdot X_{\mathrm{Best}} + a_3 \cdot (X_{\mathrm{r1}} - X_{\mathrm{r2}}) \quad (24)$$

In Eq. (24), $a_1$, $a_2$, and $a_3$ are three random numbers between 0 and 1, satisfying the constraint $a_1 + a_2 + a_3 = 1$. After creating the new individual *V*, a selection operation is performed: if individual *V* is superior to individual $X_{\mathrm{k}}$, then *V* replaces $X_{\mathrm{k}}$; otherwise, *V* is discarded.

The global neighborhood search-based adaptive differential evolution algorithm enhances its performance by integrating the global neighborhood search strategy. This strategy utilizes the information of the current optimal solution to guide the search direction, significantly improving local exploitation capabilities. Meanwhile, the random difference vector prevents premature convergence, making it suitable for complex optimization scenarios in TAB multi-port transmission.

## IV. SIMULATION VERIFICATION

According to the proposed IEPSM strategy, simulations are conducted under three different operating conditions to observe the current ripple of the system under both the optimal phase shift strategy and the existing ripple optimization control scheme at the same output power. The parameters are shown in Table III. This study conducted verification in a dual-input single-output configuration, comparing three strategies—SPS, IPS [24], and the proposed IEPSM—under three different output current settings to evaluate their performance in suppressing output current ripple.

TABLE III
TRIPLE ACTIVE BRIDGE SYSTEM PARAMETER

| Parameter name | Specific values |
|---|---|
| Switching cycle T | 100μs |
| Port 1 voltage amplitude $U_1$ | 144V |
| Port 2 voltage amplitude $U_2$ | 50V |
| 1-port inductance $L_1$ | 8μH |
| 2-port inductance value $L_2$ | 8μH |
| 3-port inductance $L_3$ | 1μH |
| $E_{\mathrm{rev}}$ | 4.38V |
| $R_{\mathrm{a}}$ | 0.035 Ω |
| $R_{\mathrm{mem}}$ | 0.088 Ω |
| $C_{\mathrm{a}}$ | 37.26 F |
| $C_{\mathrm{c}}$ | 37.26 F |

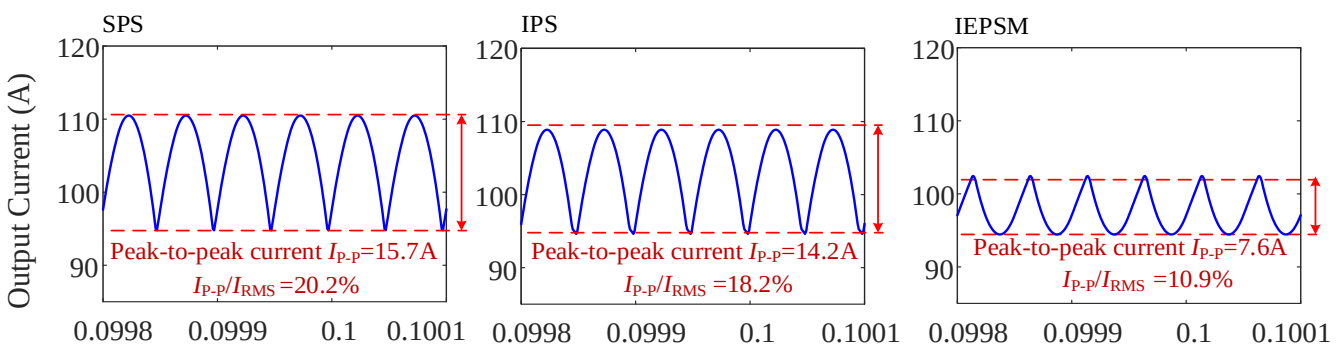

Fig. 10. Output current ripple under SPS, IPS, and IEPSM ($I_{\mathrm{out}}$=100A).

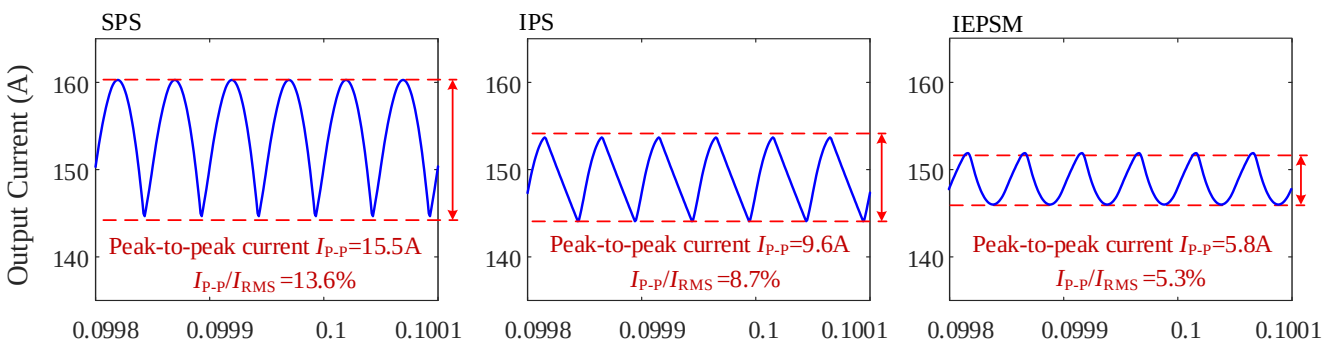

Fig. 11. Output current ripple under SPS, IPS, and IEPSM ($I_{\mathrm{out}}$=150A).

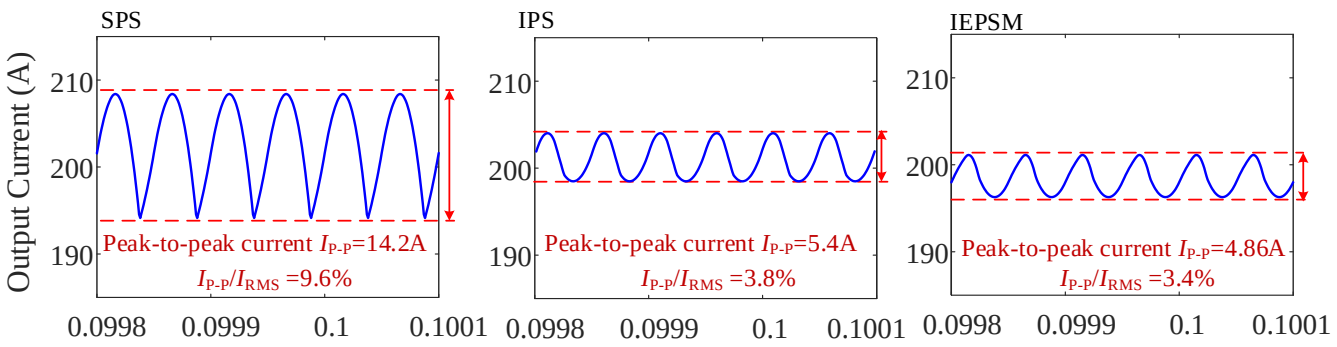

Fig. 12. Output current ripple under SPS, IPS, and IEPSM ($I_{\mathrm{out}}$=200A).

Figs. 10, 11, and 12 respectively show the comparison results of three different output current ripple suppression strategies when the output current is set to 100 A, 150 A, and 200 A. Taking 100 A as an example, the modulation strategies from left to right are SPS, IPS, and IEPSM. The output current ripple of the SPS strategy is 15.7 A, with a ripple rate of 15.7%; the output current ripple of the IPS strategy is 14.2 A, with a ripple rate of 14.2%; and the output current ripple of the IEPSM

strategy is 7.1 A, with a ripple rate of 7.1%. Furthermore, as shown in Figs. 10 and 11, the IEPSM strategy effectively suppresses the current ripple under different loads and achieves better performance than the traditional SPS and IPS strategies.

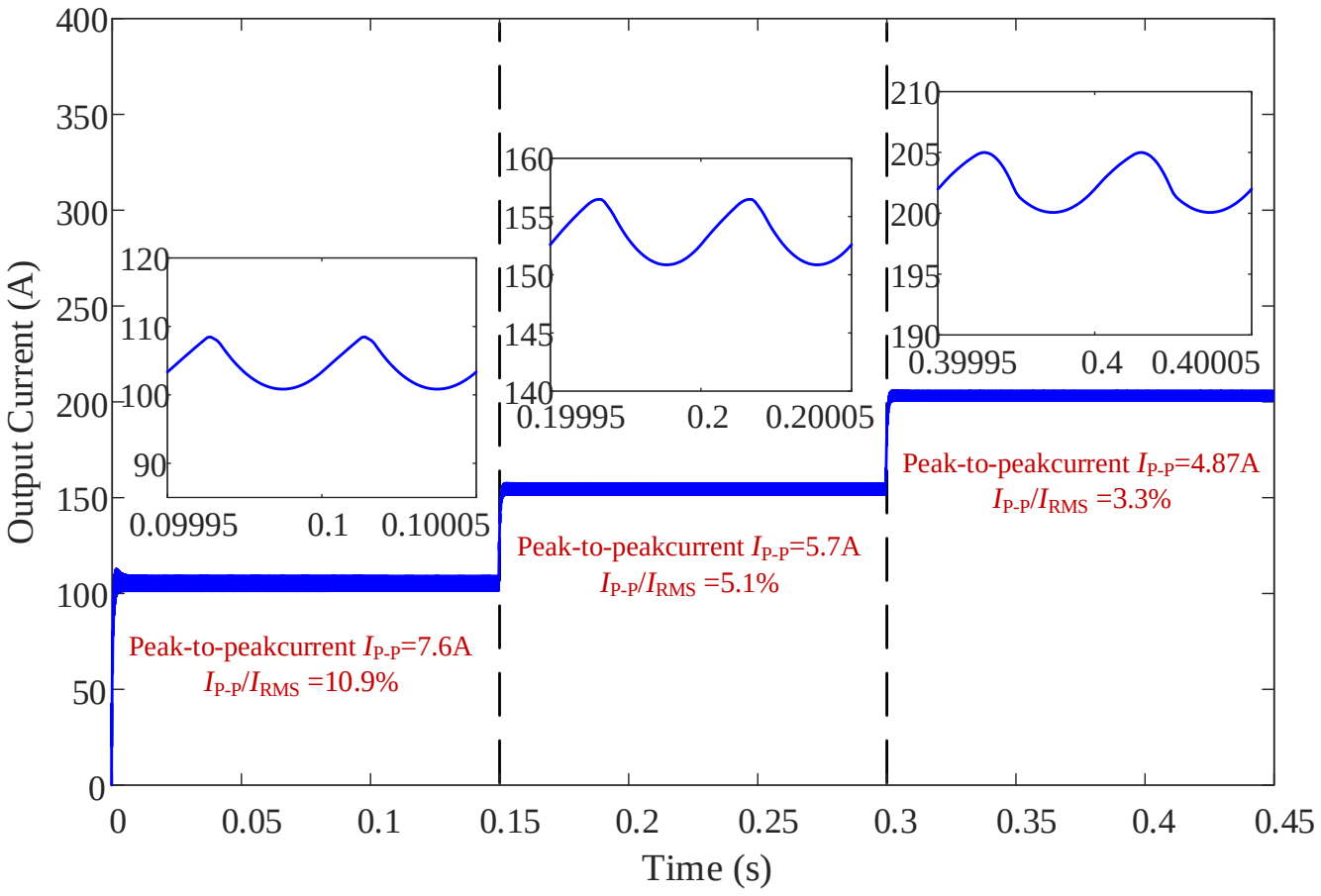

Fig. 13. Current ripple optimization under different output current conditions ($I_{out}$=100/150/200A).

The system sets the output current starting from 100A and increasing to 150A and 200A every 0.15 seconds. The experimental results demonstrate that IEPSM dynamically reduces ripple under changing electrolyzer loads. As shown in Fig. 13, the waveform quickly adapts to changes and matches the steady-state performance, verifying the dynamic response capability of the IEPSM strategy.

## V. Experimental Results

In order to verify the IEPSM strategy proposed in this paper, an experimental verification is carried out. In this work, a hardware-in-the-loop (HiL) platform for the triple-port TAB system was developed using the OP5707 real-time simulator on the RT-LAB platform. Port 3's load is modeled as a PEM electrolyzer. The experimental platform is shown in Fig. 14, and the platform parameters are shown in Table III.

Figs. 15 and 16 present the comparison results of different strategies under varying output current conditions. $CH_1$ represents the input voltage u1 at port 1, $CH_2$ denotes the output voltage $u_{out}$, $CH_3$ indicates the output current $I_{out}$, and $CH_4$ shows the output current $I_3$ at port 3.

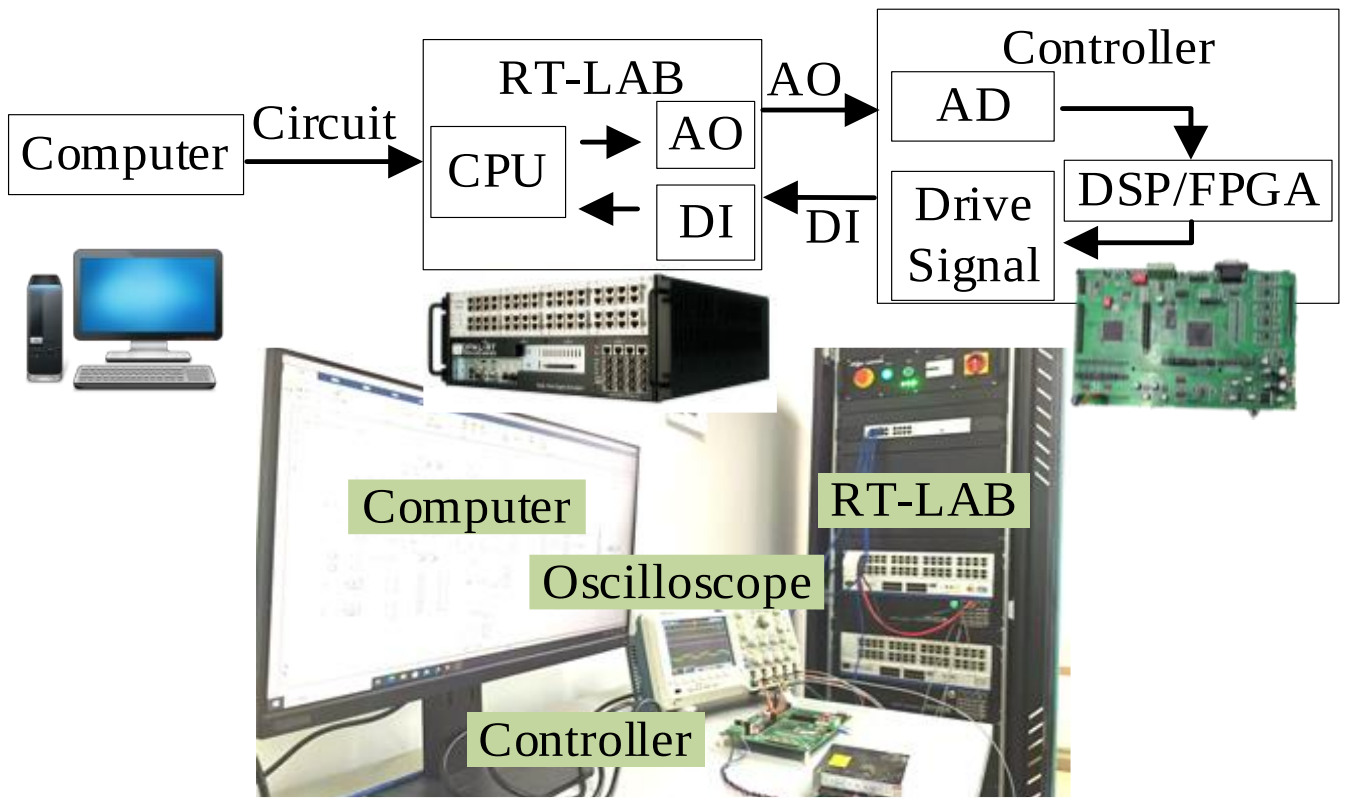

Fig.14: Experimental platform.

In the offline stage, experimental data of key variables (voltage, current, and load conditions) is collected to train the DE algorithm and obtain optimal hyperparameters, which are embedded into the FPGA and DSP controller. In the online stage, real-time parameters such as $I_{out}$ are acquired, filtered, and processed to calculate the minimum $I_{peak}$ and corresponding phase-shift angles, which are dynamically adjusted to achieve continuous current ripple suppression.

As shown in Fig. 15(a), the input voltage at port 1 is 144V, the output current is 150A, and the peak inductor current at port 3 is 600A. Theoretical calculations and experimental results indicate that at the same set output current level, different strategies yield different current peaks, and the current ripple varies accordingly. In Fig. 15(d), the peak-to-peak output current is 16.4A; in Fig. 15(e), it is 9.6A; and in Fig. 15(f), it is around 5.8A. The experimental results show that under the IEPSM strategy, the peak-to-peak value decreases to 35.4% of that observed under the SPS strategy and to 60.4% of that observed under the IPS strategy.

Fig. 16 presents the experimental waveforms for an output current of 200A. In Fig. 16(d), the peak-to-peak output current is 15.1A. In Fig. 16(e), the peak-to-peak output current is 6.2A, while in Fig. 16(f), it is 4.86A.

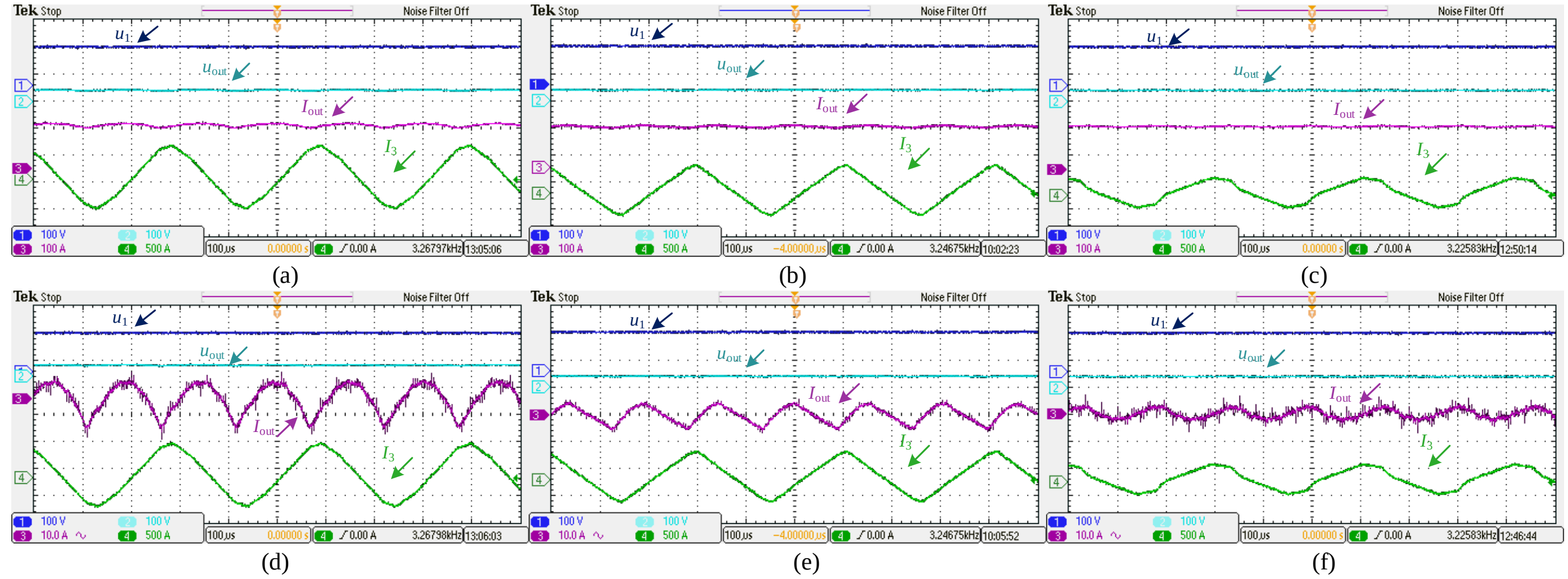

Fig. 15. Comparison of output current waveforms under different phase-shift strategies ($I_{out}$ = 150 A). (a) SPS strategy. (b) IPS strategy. (c) IEPSM strategy. (d) SPS strategy (enlarged). (e) IPS strategy (enlarged). (f) IEPSM strategy (enlarged).

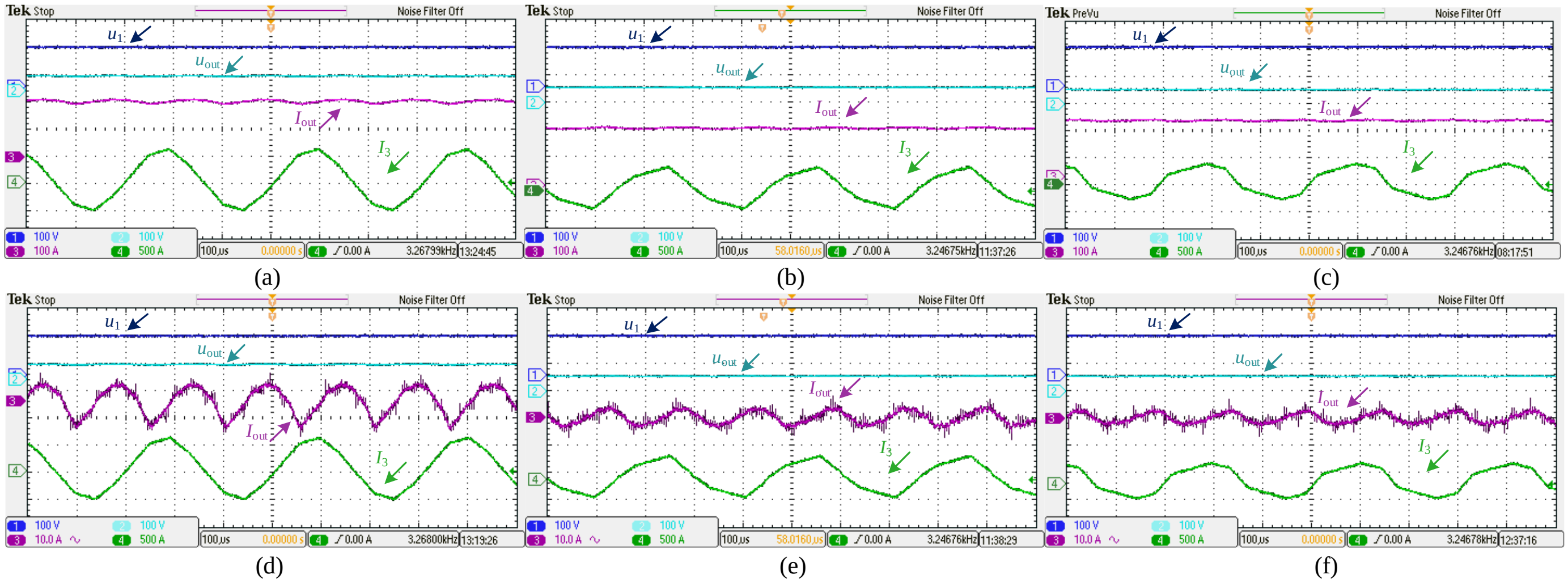

Fig. 16. Comparison of output current waveforms under different phase-shift strategies ($I_{out}$ = 200 A). (a) SPS strategy. (b) IPS strategy. (c) IEPSM strategy. (d) SPS strategy (enlarged). (e) IPS strategy (enlarged). (f) IEPSM strategy (enlarged).

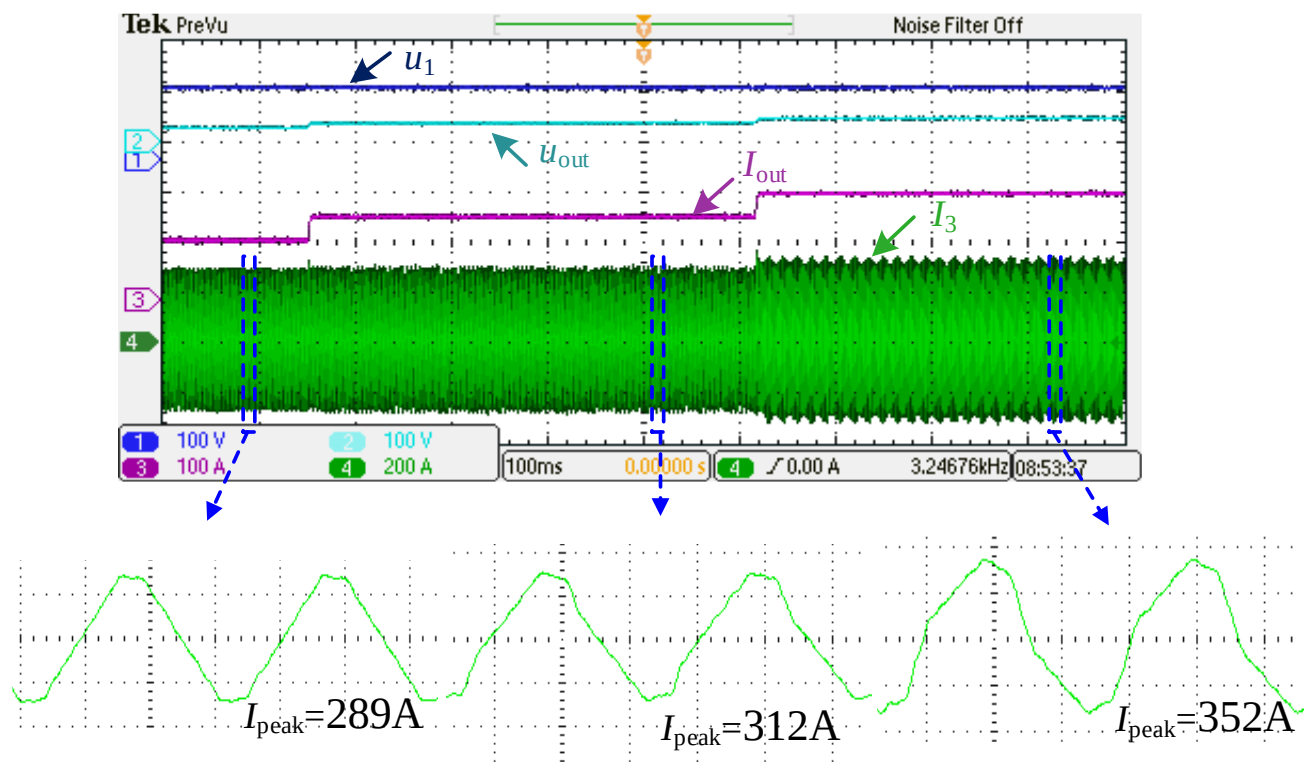

Fig. 17. Continuous dynamic response of the system under different output current levels ($I_{out}$ = 100A, 150A, 200A).

The experimental results in Figs. 15 and 16 demonstrate that the proposed IEPSM strategy in this paper can meet the system's output requirements. Compared to the SPS and IPS strategies, the output current ripple has been optimized. In the photoelectrochemical hydrogen production system, the proposed scheme can improve hydrogen production efficiency by reducing current ripple.

Fig. 17 shows the variation of the output current as the set value changes from 100A to 150A and then to 200A, simulating the fluctuation of the electrolyzer load. By dynamically optimizing with the improved differential evolution algorithm using adaptive parameters, the optimal phase-shift angle combination is rapidly obtained. The output current ripple is dynamically optimized to the minimum value at different operating points. This verifies the dynamic performance of the proposed IEPSM strategy.

TABLE IV
ALGORITHM PARAMETERS

| Objective Function | Minimum $I_{peak}$ value for a given $I_{out}$ |
|---|---|
| Population size | 500 |
| Maximum number of iterations | 500 |
| Initial zoom factor range | 0.6-1 |
| Initial crossover probability range | 0.8-1 |

This paper compares the improved DE algorithm, the traditional DE algorithm, the Genetic Algorithm (GA), and Particle Swarm Optimization (PSO) under the IEPSM strategy. To ensure a fair comparison, the circuit parameters, population size, and maximum iteration count are kept consistent across all four algorithms.

Following the flowchart shown in Fig. 9, this paper employs an improved differential evolution algorithm based on an adaptive parameter strategy to search for the optimal phase shift combination under the IEPSM strategy. The key parameters of the algorithm are listed in TABLE IV. Based on the optimization results shown in Fig. 18, it can be observed that under different electrolyzer load conditions, the improved DE algorithm exhibits superior stability compared to the other three optimization algorithms, including PSO, GA, and the conventional DE algorithm. The peak current optimization results of the triple-port system obtained by the improved DE algorithm are consistently better, demonstrating its effectiveness in minimizing current peaks under various load conditions.

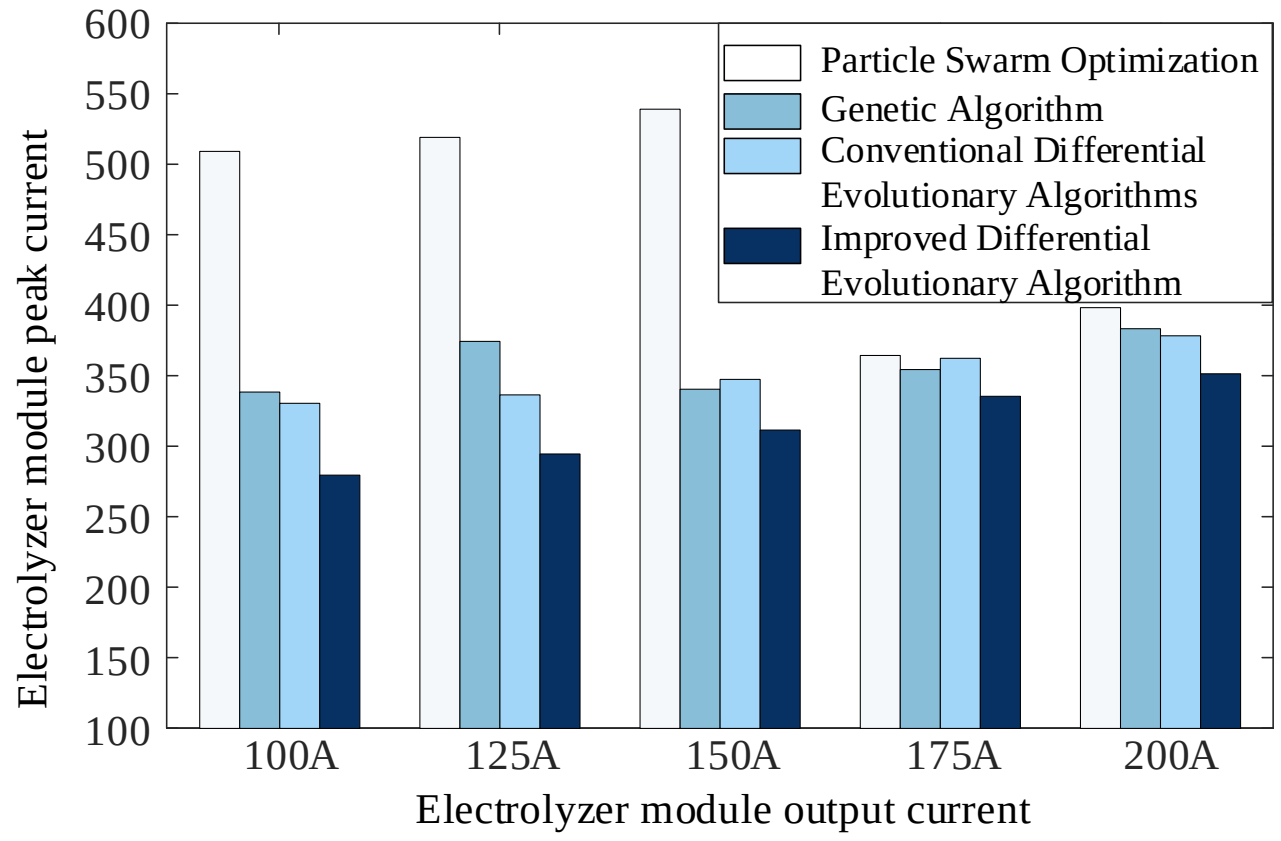

Fig. 18. The peak current under different optimization algorithms ($I_{out}$=100, 125, 150, 175, 200A).

As illustrated in Fig. 19, the comparison of output current ripple under different optimization strategies (SPS, IPS and IEPSM) shows that the IEPSM strategy can significantly reduce the current ripple. When the output current is 200A, the ripple is reduced by 67.8% compared with the SPS strategy and by 21.6% compared with the IPS strategy.

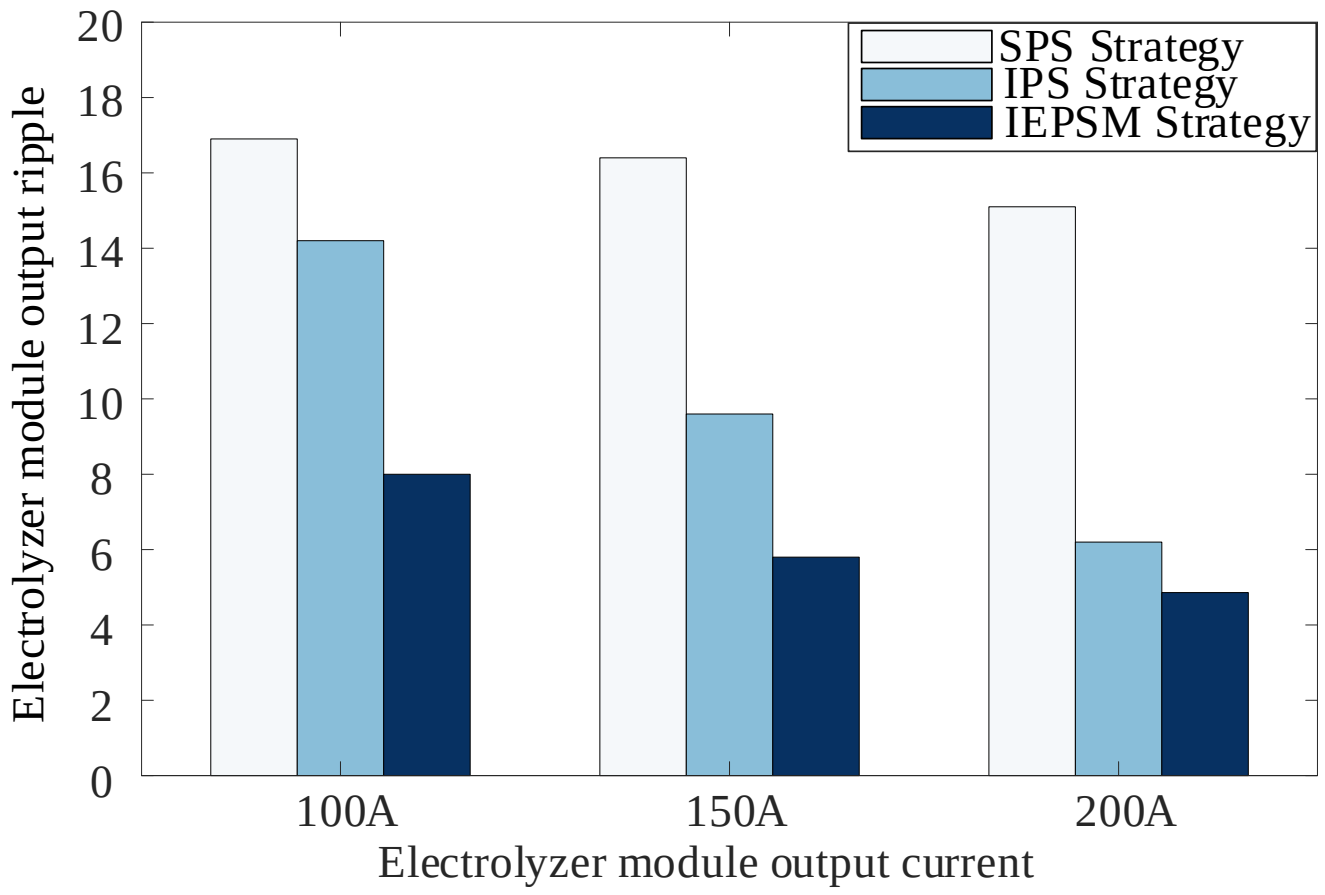

Fig. 19. Comparison of output current ripple magnitude under different modulation strategies.

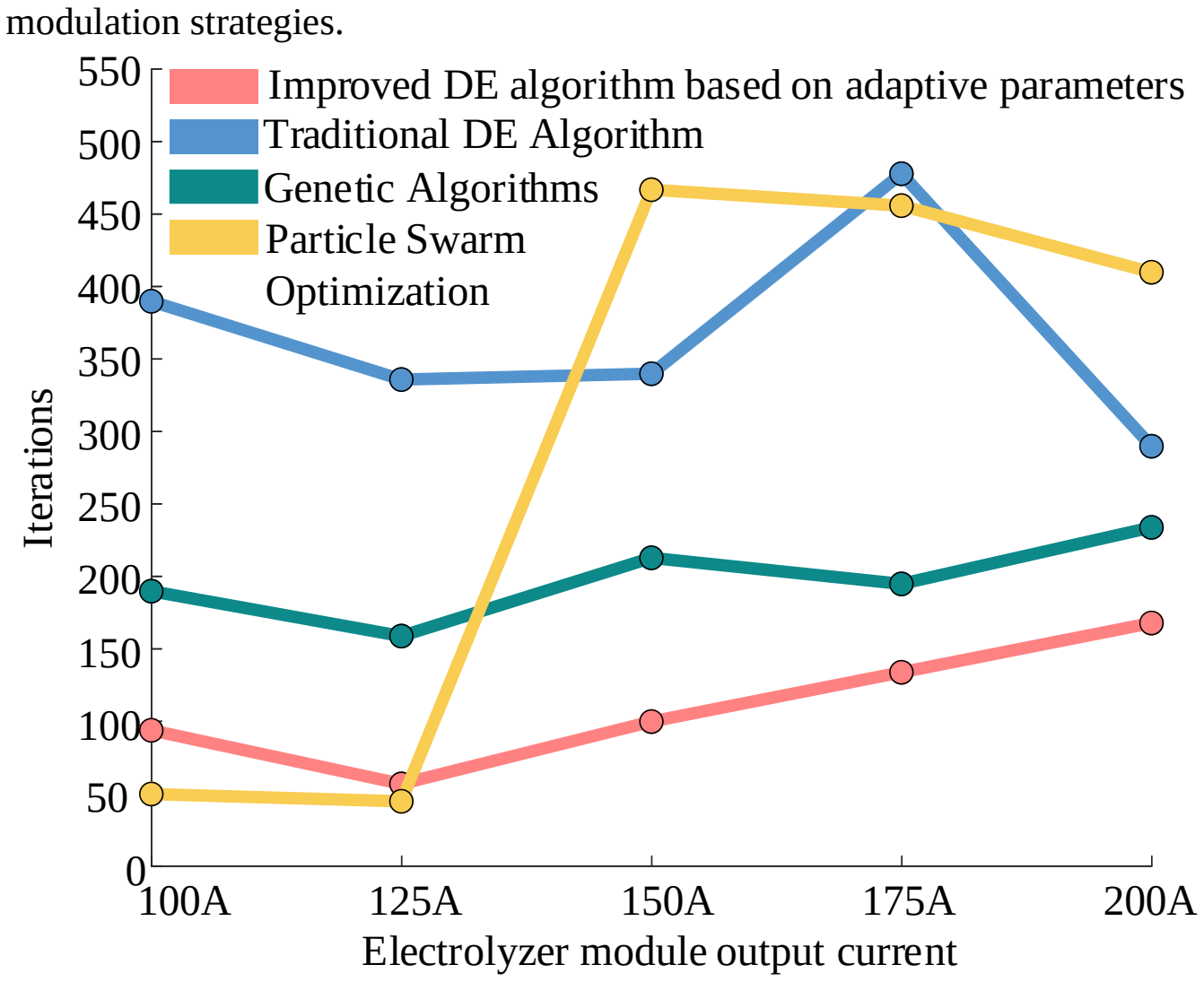

Fig. 20. Comparison of iteration counts under different optimization algorithms.

As shown in Fig. 20, when the Electrolyzer module output current is 100A and 125A, the improved DE algorithm and the PSO algorithm exhibit similar iteration counts. However, the peak current of the improved DE algorithm achieves approximately 280A when the Electrolyzer module output current is 100A and 295A when the Electrolyzer module output current is 125A. Compared with the PSO algorithm, the peak current of the improved DE algorithm is reduced by 45.1% and 43.3% respectively. In contrast, the genetic algorithm and the traditional DE algorithm require more iterations and result in higher peak currents. When the Electrolyzer module output current is 150A, 175A, and 200A, the improved DE algorithm and the PSO algorithm yield phase-shift combinations with similar peak currents, but the improved DE algorithm converges faster with fewer iterations. The reduction in the number of iterations directly benefits real-time control applications by lowering the online computation cost and accelerating the system response. Again, the genetic algorithm and the traditional DE algorithm demonstrate higher iteration counts and larger peak currents. Across the Electrolyzer module output current range from 100A to 200A, the improved DE algorithm consistently requires fewer iterations and calculates phase-shift combinations with smaller peak currents, validating the effectiveness of the proposed algorithm. From Fig. 21, it is evident that the application of the global neighborhood search strategy accelerates convergence, while the adaptive parameter control and elite retention strategies enhance the algorithm's exploitation capability, enabling it to seek global optimal solutions.

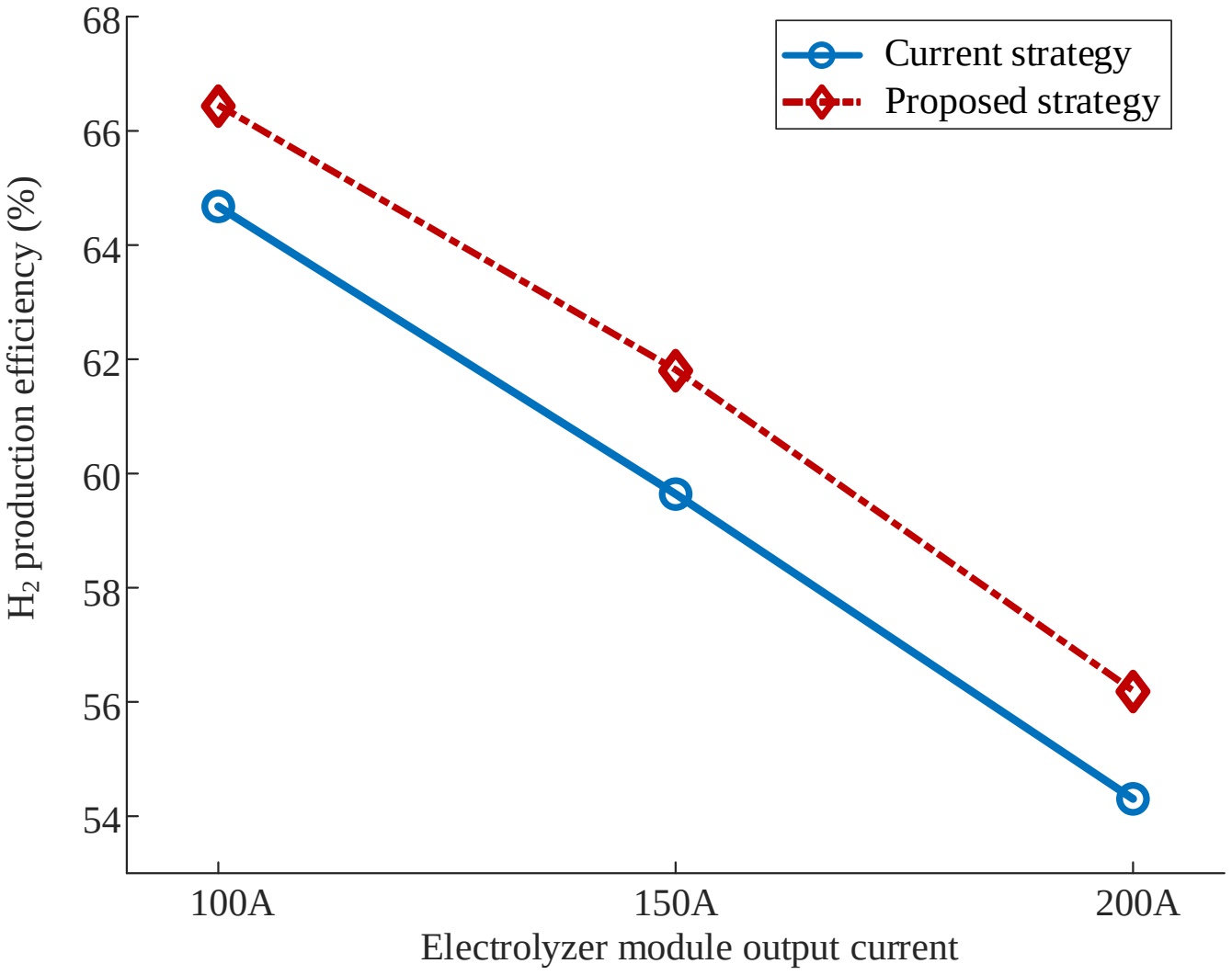

Fig. 21. Comparison of hydrogen production efficiency between the existing strategy and the proposed strategy.

This paper conducts a quantitative analysis of the electrolyzer efficiency. The hydrogen production efficiency model of the electrolyzer is expressed as:

$$\eta_{el} = \frac{n_{H_2} \cdot LHV}{P_{el}} = \frac{n_{H_2} \cdot LHV}{U \cdot I} \tag{25}$$

$$U = U_{rev} + U_{ohm} + U_{act} \tag{26}$$

where $n_{H2}$ represents the hydrogen production rate, $LHV$ is the lower heating value of hydrogen, and $P_{el}$ is the power consumption of the electrolyzer. The electrolyzer voltage $U$ is composed of three parts: the reversible voltage $U_{rev}$, the ohmic overvoltage $U_{ohm}$, and the activation overvoltage $U_{act}$. The electrolysis current is denoted as $I$. However, this model does not consider the influence of current ripple, which may affect the performance and durability of the system under fluctuating operating conditions.

Current ripple leads to an increase in both ohmic losses and activation losses. It also increases the effective value of the electrolysis current[18, 19]. Therefore, the hydrogen production efficiency is calculated as follows:

$$\eta_{el,\ ripple} = \frac{n_{H_2} \cdot LHV}{(U_{rev} + U_{ohm} + U_{act})\sqrt{I_{DC}^2 + I_{AC}^2}} \tag{27}$$

Therefore, based on Eq. (27) , and referring to the PEM electrolyzer efficiency model [36], combined with the experimental results shown in Fig. 19, the hydrogen production efficiency of the electrolyzer is obtained and illustrated. As shown in Fig. 21, the current strategy is implemented based on the SPS method.

## VI. Conclusion

This paper presented a low-ripple modulation strategy based on inner and outer phase-shift angle optimization (IEPSM) for a photovoltaic-based triple-port hydrogen production system. The proposed method was thoroughly analyzed and validated. The major contributions are summarized as follows:

1) IEPSM Strategy and Analytical Modeling: We introduced the IEPSM strategy to increase the system's control flexibility by independently adjusting inner and outer phase shifts. Analytical expressions for the hydrogen production port's output current and its peak value under IEPSM were derived, providing a precise basis for predicting and minimizing current ripple.

2) Improved DE Algorithm for Optimization: A precise mathematical model of the triple-port converter was developed, and an improved Differential Evolution (DE) algorithm with an adaptive parameter strategy was implemented to optimize the phase-shift angles. This enhanced DE algorithm significantly accelerates convergence and improves computational accuracy. In our simulations, it consistently required fewer iterations (up to about 30–40% fewer at higher load currents) to find the global optimum compared to the standard DE and genetic algorithm, while achieving lower peak currents.

3) Simulation and Experimental Validation of Ripple Reduction: Extensive simulation and experimental results demonstrate that the proposed IEPSM scheme achieves superior ripple reduction compared to conventional single-phase shift (SPS) and inner-phase shift (IPS) strategies. For example, at a 150 A output, the IEPSM strategy reduced the peak-to-peak output current ripple from approximately 16.4 A (under SPS control) to about 5.8 A, which is only 35.4% of the baseline ripple (a 64.6% reduction). It also lowered the ripple to 60.4% of that under the IPS strategy (around a 39.6% reduction). Even under dynamic load fluctuations (with the output current stepping from 100 A to 150 A to 200 A), the IEPSM approach rapidly recalculated the optimal phase-shift combination, minimizing the output current ripple at each operating point. This robust performance under varying conditions highlights the effectiveness of the proposed strategy in achieving optimal ripple performance during load transients.

The proposed IEPSM-based modulation and optimization approach achieves dynamic ripple minimization in a triple-port photovoltaic electrolysis hydrogen production system. By dynamically optimizing and effectively reducing the ripple characteristics of the electrolyzer's operating current, the proposed strategy can help enhance both its efficiency and operational longevity. In addition, its adaptability to fluctuating input makes it well-suited for renewable-powered hydrogen production systems, supporting stable operation under dynamic conditions. Future work will explore the integration of advanced intelligent optimization algorithms and the extension of the proposed low-ripple control strategy to multi-energy complementary hydrogen production systems, aiming to improve system-level performance and broaden its practical engineering applicability.

## References

[1] R. Takahashi, H. Kinoshita, T. Murata *et al.*, "Output Power Smoothing and Hydrogen Production by Using Variable Speed Wind Generators," *IEEE Transactions on Industrial Electronics,* vol. 57, no. 2, pp. 485-493, Feb. 2010.

[2] Q. Song, X. Chen, J. Liu *et al.*, "Decentralized Energy Management of Renewable Hydrogen Production Systems Without Communication Networks," *IEEE Transactions on Industrial Electronics,* vol. 71, no. 8, pp. 8927-8937, Aug. 2024.

[3] L. Zhang, Y. Qiu, Y. Chen, and A. T. Hoang, "Multi-objective particle swarm optimization applied to a solar-geothermal system for electricity and hydrogen production; Utilization of zeotropic mixtures for performance improvement," *Process Safety and Environmental Protection,* vol. 175, pp. 814-833, Jul. 2023.

[4] C. Zou, J. Li, X. Zhang *et al.*, "Industrial status, technological progress, challenges, and prospects of hydrogen energy," *Natural Gas Industry B,* vol. 9, no. 5, pp. 427-447, Oct. 2022.

[5] M. Yue, H. Lambert, E. Pahon *et al.*, "Hydrogen energy systems: A critical review of technologies, applications, trends and challenges," *Renewable and Sustainable Energy Reviews,* vol. 146, pp. 111180, Aug. 2021.

[6] X. Meng, M. Chen, M. He *et al.*, "A novel high power hybrid rectifier with low cost and high grid current quality for improved efficiency of electrolytic hydrogen production," *IEEE transactions on power electronics,* vol. 37, no. 4, pp. 3763-3768, Apr. 2022.

[7] M. El-Shafie, "Hydrogen production by water electrolysis technologies: A review," *Results in Engineering,* vol. 20, pp. 101426, Dec. 2023.

[8] O. E. Oyewole, A. A. Abdelaziz, I. A. Jimoh *et al.*, "Optimised linear active disturbance rejection control of multiport-isolated DC-DC converter for hydrogen energy storage system integration," *Alexandria Engineering Journal,* vol. 102, pp. 159-168, Sept. 2024.

[9] A. Hassan, O. Abdel-Rahim, M. Bajaj, and I. Zaitsev, "Power electronics for green hydrogen generation with focus on methods, topologies, and comparative analysis," *Scientific Reports,* vol. 14, no. 1, pp. 24767, Oct. 2024.

[10] L. Wang, H. Wang, M. Fu *et al.*, "A Three-Port Energy Router for Grid-Tied PV Generation Systems With Optimized Control Methods," *IEEE Transactions on Power Electronics,* vol. 38, no. 1, pp. 1218-1231, Jan. 2023.

[11] L. Jiang, J. Qu, L. Kong *et al.*, "Novel integrated three-port DC/DC converter for different operating modes between renewable energy, electrolyser and fuel cell/battery," *Clean Energy,* vol. 7, no. 1, pp. 174-189, Mar. 2023.

[12] L. Xu, C. Dou, D. Yue *et al.*, "Hybrid Modeling and Switching Control of Electric Vehicle Aggregation for Frequency Regulation," *IEEE Transactions on Sustainable Energy*, pp. 1-14. 2025.

[13] F. Wu, F. Feng, and H. B. Gooi, "Cooperative Triple-Phase-Shift Control for Isolated DAB DC–DC Converter to Improve Current Characteristics," *IEEE Transactions on Industrial Electronics,* vol. 66, no. 9, pp. 7022-7031, Sept. 2019.

[14] F. Lin, X. Zhang, X. Li *et al.*, "Automatic Triple Phase-Shift Modulation for DAB Converter With Minimized Power Loss," *IEEE Transactions on Industry Applications,* vol. 58, no. 3, pp. 3840-3851, May-Jun. 2022.

[15] Z. Guo, M. Li, and X. Han, "Triple-Phase Shift Modulation Scheme of DAB Converter With LCL Resonant Tank," *IEEE Transactions on Transportation Electrification,* vol. 8, no. 2, pp. 1734-1747, Jun. 2022.

[16] A. Tello, F. A. Cataño, A. Arunachalam *et al.*, "Green hydrogen production by photovoltaic-assisted alkaline water electrolysis: A review on the conceptualization and advancements," *International Journal of Hydrogen Energy,* vol. 107, pp. 378-395, Mar. 2025.

[17] X. Guo, H. Zhu, and S. Zhang, "Overview of electrolyser and hydrogen production power supply from industrial perspective," *International Journal of Hydrogen Energy,* vol. 49, pp. 1048-1059, Jan. 2024.

[18] V. Ruuskanen, J. Koponen, A. Kosonen *et al.*, "Power quality estimation of water electrolyzers based on current and voltage measurements," *Journal of Power Sources,* vol. 450, pp. 227603, Feb. 2020.

[19] H. P. C. Buitendach, R. Gouws, C. A. Martinson *et al.*, "Effect of a ripple current on the efficiency of a PEM electrolyser," *Results in Engineering,* vol. 10, pp. 100216, Jun. 2021.

[20] F. Parache, H. Schneider, C. Turpin *et al.*, "Impact of Power Converter Current Ripple on the Degradation of PEM Electrolyzer Performances," *Membranes,* vol. 12, no. 2, pp. 109, Jan. 2022.

[21] H. Bai, and C. Mi, "Eliminate Reactive Power and Increase System Efficiency of Isolated Bidirectional Dual-Active-Bridge DC–DC

Converters Using Novel Dual-Phase-Shift Control," *IEEE Transactions on Power Electronics*, vol. 23, no. 6, pp. 2905-2914, Nov. 2008.
[22] X. Liu, Z. Q. Zhu, D. A. Stone *et al.*, "Novel Dual-Phase-Shift Control With Bidirectional Inner Phase Shifts for a Dual-Active-Bridge Converter Having Low Surge Current and Stable Power Control," *IEEE Transactions on Power Electronics*, vol. 32, no. 5, pp. 4095-4106, May. 2017.
[23] Y. Xu, T. Liu, Y. Chen *et al.*, "Intrinsically safe DAB converter for reliable power supply in harsh environment via extended phase shift pulse train control," *IET Power Electronics*, vol. 17, no. 16, pp. 2808-2818, Sept. 2024.
[24] X. Guo, H. Zhu, H. Zhang *et al.*, "A Current Ripple Suppression Strategy of TAB for Photovoltaic Hydrogen Production," *IEEE Transactions on Industrial Electronics*, vol. 72, no. 4, pp. 3758-3767, Apr. 2025.
[25] Á. Hernández-Gómez, V. Ramirez, and D. Guilbert, "Investigation of PEM electrolyzer modeling: Electrical domain, efficiency, and specific energy consumption," *International Journal of Hydrogen Energy*, vol. 45, no. 29, pp. 14625-14639, May. 2020.
[26] X. Guo, S. Zhang, Y. Gao *et al.*, "Advancements in Photovoltaic Electrolysis for Green Hydrogen Production: A Comprehensive Review and Comparative Analysis of Modeling Approaches," *IEEE Transactions on Power Electronics*, vol. 40, no. 7, pp. 10000-10026, Jul. 2025.
[27] X. Meng, Q. Duan, G. Sha *et al.*, "An Efficiency Improvement Strategy for Triple-Active-Bridge-Based DC Energy Routers in DC Microgrids," *Electronics*, vol. 13, no. 7, pp. 1172, Feb. 2024.
[28] D. Guilbert, S. M. Collura, and A. Scipioni, "DC/DC converter topologies for electrolyzers: State-of-the-art and remaining key issues," *International Journal of Hydrogen Energy*, vol. 42, no. 38, pp. 23966-23985, Sept. 2017.
[29] Z. Dobó, and Á. B. Palotás, "Impact of the current fluctuation on the efficiency of Alkaline Water Electrolysis," *International Journal of Hydrogen Energy*, vol. 42, no. 9, pp. 5649-5656, Mar. 2017.
[30] R. Storn, and K. Price, "Differential evolution–a simple and efficient heuristic for global optimization over continuous spaces," *Journal of global optimization*, vol. 11, pp. 341-359, Dec. 1997.
[31] L. Xu, C. Dou, D. Yue *et al.*, "End-Edge-Cloud Collaboration-Based EVs Aggregator Control Method for Multiple Frequency Regulation Considering User Charging Demand," *IEEE Transactions on Transportation Electrification*, vol. 11, no. 1, pp. 5017-5028, Feb. 2025.
[32] A. W. Mohamed, A. A. Hadi, and A. K. Mohamed, "Differential Evolution Mutations: Taxonomy, Comparison and Convergence Analysis," *IEEE Access*, vol. 9, pp. 68629-68662, May. 2021.
[33] Y. Zhou, X. Li, and L. Gao, "A differential evolution algorithm with intersect mutation operator," *Applied Soft Computing*, vol. 13, no. 1, pp. 390-401, Jan. 2013.
[34] Z. Luo, X. Qian, and W. Song, "Enhanced differential evolution with hierarchical selection mutation and distance-based selection strategy," *Engineering Applications of Artificial Intelligence*, vol. 144, pp. 110124, Mar. 2025.
[35] Y. L. Li, Z. H. Zhan, Y. J. Gong *et al.*, "Differential Evolution with an Evolution Path: A DEEP Evolutionary Algorithm," *IEEE Transactions on Cybernetics*, vol. 45, no. 9, pp. 1798-1810, Sept. 2015.
[36] M. Abomazid, N. A. El-Taweel and H. E. Z. Farag, "Novel Analytical Approach for Parameters Identification of PEM Electrolyzer," *IEEE Transactions on Industrial Informatics*, vol. 18, no. 9, pp. 5870-5881, Sept. 2022.

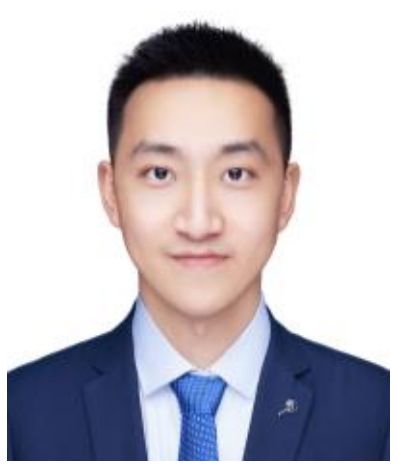

**Shiqi Zhang** (Graduate Student member, IEEE) received M.S.degrees in emerging power system from Curtin University, Perth, Australia, in 2021. He is pursuing a Ph.D. at Yanshan University, Qinhuangdao, China.

He has authored/coauthored twenty-five patents and fifteen technical papers. His research interests include hydrogen production converters.

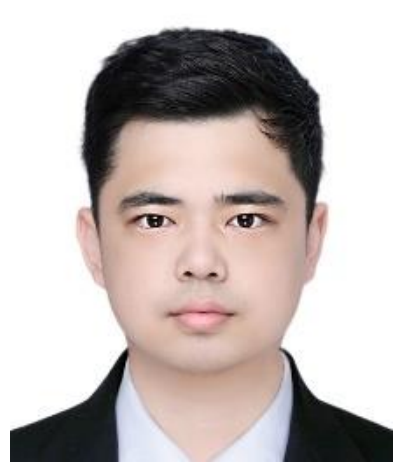

**Ziang Jiao** was born in Shijiazhuang, China. He received the B.S degree in electrical engineering from Hebei University of Technology, China in 2023. He is currently working toward an His current research interests include: design of three-port hydrogen production converter and design and control of power electronic converters.

M. degree in electrical engineering from the University of Yanshan, China.

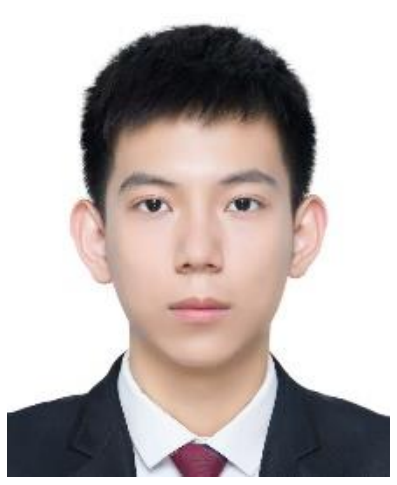

**Jiaxin Su** was born in Shijiazhuang, China. He is pursuing a degree in electrical engineering in Qinhuangdao, Hebei Province, in 2023.

He has accepted/authorized two national invention patents and three soft papers. His research focuses on hydrogen generation converters and the application of power electronics to power systems.

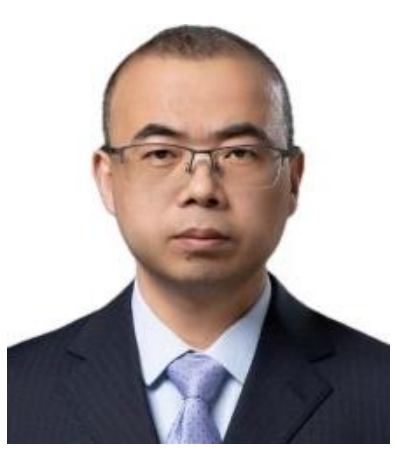

**Ning Wang** was born in January 1981 in Hebei, China. He received his Bachelor's and Master's degrees from Yanshan University in 2003 and 2006, respectively, and his PhD from Tianjin University in 2013. He is currently an associate professor at Yanshan University. He is parti-cularly interested in the operation control of electric hydrogen systems and new energy generation technology.

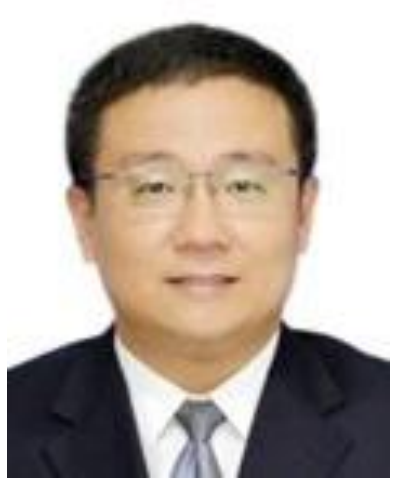

**Zheng Li** (Member,IEEE) was born in Shijiazhuang, China, in 1980. He received the B.Sc. and Ph.D. degrees in electrical engineering and power electronics and electric drive from Hefei University of Technol ogy, Hefei, China, in 2002 and 2007, respectively. Since 2007, he has been a Lecturer, Associate Professor, and Professor with the School of Electrical Engineering, Hebei University of Science and Technology. From July 2013 to July 2014, he has been a visiting scholar and part-time faculty with the College of Engineering, Wayne State University, USA. He has authored more than 400 published papers. His current research interests include renewable energy applications, design, analysis, and control of novel motors and actuators, intelligent control, and power electronics.

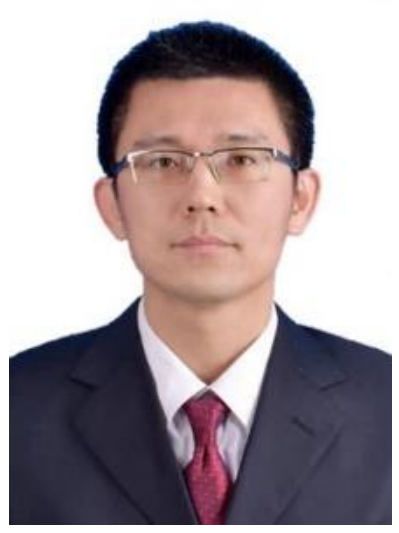

**Xiaoqiang Guo** (Senior Member, IEEE) re-ceived the B.S. and Ph.D. degrees in electrical engineering from Yanshan University, Qinhuangdao, China, in 2003 and 2009, respectively.

He has been a Postdoctoral Fellow with the Laboratory for Electrical Drive Applications and Research (LEDAR), Ryerson University, Toronto, ON, Canada. He is currently a professor with the Department of Electrical Engineering, Yanshan University, China. He has authored/coauthored more than eighty technical papers, in addition to eleven patents. His current research interests include high-power converters and ac drives, electric vehicle charging station, and renewable energy power conversion systems.

Dr. Guo is a Senior Member of the IEEE Power Electronics Society and IEEE Industrial Electronics Society. He is currently an Associate Editor of the IET Power Electronics, Journal of Power Electronics, and CPSS Transactions on Power Electronics and Applications.

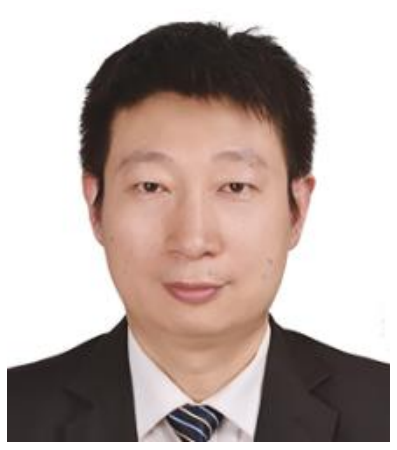

**Changchun Hua** (Fellow, lEEE) re-ceived the Ph.D. degree in electrical engineering from Yanshan University, Qinhuangdao, China, in 2005.

He was a Research Fellow with the National University of Singapore, Singapore, from 2006 to 2007. From 2007 to 2009, he was with Carleton University, Ottawa, ON, Canada, funded by the Province of Ontario Ministry of Research and Innovation Program. From 2009 to 2010, he was with the University of Duisburg-Essen, Essen, Germany, funded by Alexander von Humboldt Foundation. He is currently a Full Professor with Yanshan University. He has authored or coauthored more than 80 papers in mathematical, technical journals, and conferences. He has been involved in more than 10 projects supported by the National Natural Science Foundation of China, the National Education Committee Foundation of China, and other important foundations. His current research interests include nonlinear control systems.